\documentclass[manuscript]{aastex}
\usepackage{natbib}
\usepackage{amsmath}
\usepackage{textcomp}
\usepackage[colorlinks=true,citecolor=blue]{hyperref}
\usepackage{graphicx,color}
\usepackage{times}
\usepackage{amssymb}
\newcommand{\Msun}{\ensuremath{M_{\odot}}}
\newcommand{\lum}{erg\,s$^{-1}$}
\newcommand{\fermi}{{\it Fermi}}
\newcommand{\phflux}{\mbox{${\rm \, ph \,\, cm^{-2} \, s^{-1}}$}}
\newcommand{\ergflux}{\mbox{${\rm \, erg \,\, cm^{-2} \, s^{-1}}$}}
\newcommand{\nustar}{{\it NuSTAR}}
\slugcomment{ApJ Accepted}

\shorttitle{High Redshift Blazar S5 0836+71}
\shortauthors{Vaidehi S. Paliya}

\begin{document}

\title{The High Redshift Blazar S5 0836+71: A Broadband Study}

\author{Vaidehi S. Paliya$^{1,\,2}$} 
\affil{$^1$Indian Institute of Astrophysics, Block II, Koramangala, Bangalore-560034, India}
\affil{$^2$Department of Physics, University of Calicut, Malappuram-673635, India}
\email{vaidehi@iiap.res.in}

\begin{abstract}
A broadband study of the high redshift blazar S5 0836+71 ($z$ = 2.172) is presented. Multi-frequency light curves show multiple episodes of X-ray and $\gamma$-ray flares, while optical-UV fluxes show little variations. During the GeV outburst, the highest $\gamma$-ray flux measured is (5.22 $\pm$ 1.10) $\times$ 10$^{-6}$ \phflux~in the range of 0.1$-$300 GeV, which corresponds to an isotropic $\gamma$-ray luminosity of (1.62 $\pm$ 0.44) $\times$ 10$^{50}$ \lum, thereby making this as one of the most luminous $\gamma$-ray flare ever observed from any blazar. A fast $\gamma$-ray flux rising time of $\sim$3 hours is also noticed which is probably the first measurement of hour scale variability detected from a high redshift ($z >$ 2) blazar. The various activity states of S5 0836+71 are reproduced under the assumption of single zone leptonic emission model. In all the states, the emission region is located inside the broad line region, and the optical-UV radiation is dominated by the accretion disk emission. The modeling parameters suggests the enhancement in bulk Lorentz factor as a primary cause of the $\gamma$-ray flare. The high X-ray activity with less variable $\gamma$-ray counterpart can be due to emission region to be located relatively closer to the black hole where the dominating energy density of the disk emission results in higher X-ray flux due to inverse-Compton scattering of disk photons.
\end{abstract}

\keywords{galaxies: active --- gamma rays: galaxies --- quasars: individual (S5 0836+71) --- galaxies: jets}

\section{Introduction}\label{sec:intro}
Blazars are luminous active galactic nuclei (AGN) with powerful relativistic jets aligned close to the line of sight to the observer (jet viewing angle $\theta_{\rm v} < 1/\Gamma$, where $\Gamma$ is bulk Lorentz factor). Due to their peculiar orientation, the radiation observed from blazars is dominated by the emission from relativistic jets \citep{1978PhyS...17..265B} which transport momentum and energy to large scales. Blazars include the luminous flat spectrum radio quasars (FSRQs) characterized by strong and broad optical emission lines (rest frame equivalent width $> 5$ \AA) and BL Lac objects where optical lines are weak or absent \citep{1991ApJS...76..813S,1991ApJ...374..431S}. A new classification based on the broad line region (BLR) luminosity ($L_{\rm BLR}$) in units of Eddington luminosity ($L_{\rm Edd}$) is proposed by \citet{2011MNRAS.414.2674G}. In this scheme, FSRQs are defined as sources with $L_{\rm BLR}/L_{\rm Edd} > 5 \times 10^{-4}$. In addition to that, both classes share many common properties, such as rapid flux and polarization variations \citep{1995ARA&A..33..163W,2005A&A...442...97A}, flat radio spectra ($\alpha_r~< 0.5;~\rm{S_{\nu}}~\propto~\rm{\nu}^{-\alpha}$) at GHz frequencies, and exhibit superluminal patterns at radio wavelengths \citep{2005AJ....130.1418J}.

In the widely accepted scenario of blazars, a single population of high energy electrons present in the jet, emits over the entire electromagnetic spectrum via synchrotron and inverse Compton (IC) mechanisms, the former being relevant at lower energies and the latter dominating at higher energies. Consequently, the spectral energy distribution (SED) of blazars exhibit a double hump shape, with low energy synchrotron peak lying in between infrared to X-rays and the high energy IC hump extending up to GeV/TeV $\gamma$-rays. Possible sources of seed photons for IC scattering could be internal \citep[synchrotron self Compton or SSC,][]{1981ApJ...243..700K,1985ApJ...298..114M,1989ApJ...340..181G} or external \citep[external Compton or EC,][]{1987ApJ...322..650B,1989ApJ...340..162M,1992A&A...256L..27D} to the jet. The plausible reservoir of seed photons for EC can be the accretion disk \citep{1993ApJ...416..458D,1997A&A...324..395B}, the BLR \citep{1994ApJ...421..153S,1996MNRAS.280...67G}, and the dusty torus \citep[][]{2000ApJ...545..107B}. Moreover, the presence of the high energy peak is also attributed to hadronic processes initiated by relativistic protons co-accelerated with the electrons \citep[e.g.,][]{2003APh....18..593M,2013ApJ...768...54B}. Besides non-thermal jet radiation, the thermal emission from the accretion disk is also observed in many blazars \citep[see e.g.,][]{2010MNRAS.402..497G}. Blazars are found to follow a trend, so-called `blazar sequence', with more luminous sources (generally FSRQs) being more $\gamma$-ray dominated, and have both synchrotron and IC peaks located at lower energies than their fainter (mostly BL Lac objects) analogues \citep{1998MNRAS.299..433F}. However, \citet{2012MNRAS.420.2899G} have recently pointed that such sequence could be due to selection effects.

Within the blazar population, only the most powerful objects can be detected at high redshifts and they generally belong to the FSRQ class of AGN. According to the `blazar sequence', the SED peaks of blazars shift to lower frequencies as their luminosity increases. Since at high redshift, only the most powerful objects are expected to be visible, one expects the peaks in their SEDs to shift to lower frequencies. Thus, in high redshift blazars the synchrotron hump peaks in the sub-mm region and the IC hump peaks in the $\sim$MeV band. In such sources, as the synchrotron peak shifts to the mm region, thermal emission from the accretion disk is clearly visible in the optical band \citep[e.g.,][]{2013MNRAS.433.2182S}. Accordingly, observing IC peak in the $\gamma$-ray band ($\gamma$-ray detection is the most common blazar fingerprint) is more and more difficult as redshift increases. However, in such high redshift blazars, the high energy peak can be observed in hard X-rays due to its shift from $\sim$GeV to $\sim$MeV range. In fact, the Burst Alert Telescope (BAT) onboard the {\it Swift} satellite \citep[][]{2004ApJ...611.1005G} has detected blazars up to redshift significantly higher than the \fermi-Large Area Telescope (\fermi-LAT) itself \citep[see e.g.,][]{2013ApJS..207...19B}. Moreover, these high redshift blazars generally harbor more than a billion solar mass black holes at their centers \citep[e.g.,][]{2013MNRAS.433.2182S}.

In this work, a detailed multi-wavelength study of a high redshift blazar S5 0836+71 \citep[$z$ = 2.172;][]{1993A&AS..100..395S} is presented. This source has shown multiple episodes of flaring activities since the beginning of \fermi~operation. In particular, a giant $\gamma$-ray outburst was detected by \fermi-LAT at the end of 2011 \citep{2011ATel.3831....1C}. By analyzing the publicly available observational data, an exhaustive investigation is performed to understand this peculiar $\gamma$-ray flare and the associated nuclear emission processes. Moreover, an attempt is also made to study the physical properties of S5 0836+71 during a low activity state, including the first time observations from hard X-ray focusing telescope \nustar~\citep{2013ApJ...770..103H}.

This paper is structured as follows. In Section~\ref{sec:general}, a brief collection of the main observational properties of S5 0836+71 is reported. Section~\ref{sec:data_red} is devoted to data reduction procedures followed in this work. The results are presented in Section~\ref{sec:results} and they are discussed in Section~\ref{sec:dscsn}. Conclusions are outlined in Section~\ref{sec:summary}. Throughout the work, a $\Lambda$CDM cosmology with the Hubble constant $H_0=71$~km~s$^{-1}$~Mpc$^{-1}$, $\Omega_m = 0.27$, and $\Omega_\Lambda = 0.73$ is adopted.

\section{General Properties of S5 0836+71}\label{sec:general}
The high redshift blazar S5 0836+71 is a radio-bright source \citep[$F_{\rm 5~GHz}$ = 1.68 Jy;][]{2012ApJ...744..177L} and exhibits a flat radio spectrum \citep[$\alpha = -$0.33;][]{1981A&AS...45..367K}. Detection of one-sided kilo-parsec scale radio jet is also reported \citep{1992A&A...266...93H}. Using the observations from the Very Long Baseline Interferometry (VLBI), \citet{1998A&A...334..489O} have revealed the emergence of a new superluminal knot closely associated with the bright $\gamma$-ray state. From the observations at 1.6 and 5 GHz under the VLBI Space Observatory Program \citep[VSOP;][]{1998A&A...340L..60L}, the inner structure of the jet is found to be of helical shape which is further confirmed in a recent study by \citet{2012ApJ...749...55P}.  \citet{2010ApJ...720...41A} have proposed a spine-sheath structure of the jet using the Very Long Baseline Array observations.

The optical-UV spectrum of S5 0836+71 is dominated by thermal emission from the accretion disk \citep[e.g.,][]{2010MNRAS.405..387G}. A prominent optical outburst is also reported by \citet{1993A&A...267L..23V}, wherein the source was found to be brightened by $>$ 0.5 mag. By reproducing the optical-UV spectrum with a standard accretion disk model, \citet{2010MNRAS.405..387G} have constrained the accretion disk luminosity ($L_{\rm disk}$) and black hole mass ($M_{\rm BH}$) of S5 0836+71 as 2.25 $\times$ 10$^{47}$ \lum~and 3 $\times$ 10$^{9}$ \Msun~respectively.  

Recent studies focusing on high energy X-ray properties along with the multi-wavelength modeling confirmed the blazar nature of S5 0836+71 \citep[e.g.,][]{2006A&A...453..829F,2007ApJ...669..884S,2011MNRAS.411.2137G}. It is bright in hard X-rays and included in the 70 months {\it Swift}-BAT catalog \citep{2013ApJS..207...19B}. It was detected by the Energetic Gamma Ray Experiment Telescope onboard the Compton Gamma Ray Observatory \citep{1993ApJS...86..629T} and found to be a variable emitter of $>$ 100 MeV $\gamma$-rays. Since the launch of \fermi, it is regularly monitored by LAT and included in the recently released third \fermi-LAT catalog \citep[3FGL;][]{2015arXiv150102003T}. Its $\gamma$-ray spectrum is steep, as expected from a high redshift blazar folllowing the `blazar sequence', and is modeled by a logParabola model in the 3FGL catalog. Moreover, multiple episodes of $\gamma$-ray flaring from S5 0836+71 are also reported \citep[e.g.,][]{2011ATel.3260....1C}, particularly during late 2011 \citep{2011ATel.3831....1C}.

A detailed long term multi-wavelength study of S5 0836+71 has recently been done by \citet{2013A&A...556A..71A} with a major focus on its variability characteristics. They found the source to show flux variations across the electromagnetic spectrum with the strongest variations seen in the $\gamma$-ray band. They have not found any significant correlation between $\gamma$-ray and radio fluxes and concluded that the observed radio emission can originate from components other than that produce $\gamma$-rays. A significant curvature in the flaring state $\gamma$-ray spectrum is also reported by them. However, in another study, \citet{2013EPJWC..6104003J} have proposed the emergence of a new radio knot that coincided with the $\gamma$-ray flaring period in 2011 April and they argue that the $\gamma$-ray emission region is located at a distance of $\sim$35 pc from the central black hole.  

\section{Multi-wavelength observations}\label{sec:data_red}
\subsection{{\it Fermi}-Large Area Telescope Observations}\label{subsec:fermi}
The \fermi-LAT data used in this work were collected over the first $\sim$75 months of \fermi~operation (2008 August 5 to 2014 November 15 or MJD 54,683$-$56,976). The standard data analysis procedures as mentioned in the \fermi$-$LAT documentation\footnote{http://fermi.gsfc.nasa.gov/ssc/data/analysis/documentation/} are adopted. In the energy range of 0.1$-$300 GeV, events belonging to the SOURCE class are used. Good time intervals (GTI) are generated by applying a filter expression of ``\texttt{DATA$\_$QUAL$>$0} \&\& \texttt{LAT$\_$CONFIG==1}'' and to avoid contamination from the Earth limb $\gamma$-rays, a cut of 100$^{\circ}$ is also applied on the zenith angle. 

Throughout the analysis, the unbinned likelihood method included in the pylikelihood library of {\tt Science Tools (v9r33p0)} and post-launch instrument response functions P7REP\_SOURCE\_V15 are used. Computation of the significance of the $\gamma$-ray signal is performed by means of the maximum likelihood (ML) test statistic TS =  2$\Delta \log (\mathcal{L}$) where $\mathcal{L}$ represents the likelihood function, between models with and without a point source at the position of the source of interest. All the sources lying within 10$^{\circ}$ region of interest (ROI), centered at the position of S5 0836+71 and defined in the 3FGL catalog, are included for the analysis. The model file also includes recently released galactic diffuse emission component {\tt gll\_iem\_v05\_rev1.fit} and an isotropic component {\tt iso\_source\_v05\_rev1.txt}, as background models\footnote{http://fermi.gsfc.nasa.gov/ssc/data/access/lat/BackgroundModels.html}. All the parameters except the scaling factor of the sources within the ROI are allowed to vary during the likelihood fitting. The normalization parameters of the background models are also left free to vary. In addition to that, all the sources lying between 10$^{\circ}$ to 15$^{\circ}$ are also considered in the analysis. Their spectral parameters are kept fixed to the 3FGL catalog value. A first run of the ML analysis is performed over the period of interest and all the sources with TS $<$ 25 are removed from the model. This updated model is then used for further analysis. Though S5 0836+71 is modeled by a logParabola model in the 3FGL catalog, to generate light curves, a power law (PL) model is used as the PL indices obtained from this model show smaller statistical uncertainties when compared to those obtained from complex model fits. Moreover, since the aim is to probe the shortest timescales (hence lower photon statistics), adopting a simple PL model is appropriate. For the temporal and spectral studies, the source is considered to be detected if TS $>$ 9 which corresponds to $\sim$ 3$\sigma$ detection \citep{1996ApJ...461..396M}. Bins with TS $<$ 9 are not considered. Primarily governed by uncertainty in the effective area, the measured fluxes have energy dependent systematic uncertainties of around 10\% below 100 MeV, decreasing linearly in log(E) to 5\% in the range between 316 MeV and 10 GeV and increasing linearly in log(E) up to 15\% at 1 TeV\footnote{http://fermi.gsfc.nasa.gov/ssc/data/analysis/LAT\_caveats.html}. All errors associated with the LAT data analysis are the 1$\sigma$ statistical uncertainties.

\subsection{{\it NuSTAR} Observations}\label{subsec:nustar}
S5 0836+71 was observed twice with \nustar~(PI: Fiona Harrison) each in 2013 December and 2014 January for a total elapsed time of $\sim$50 and $\sim$70 ksec respectively\footnote{Due to low Earth orbit, the net exposure time is roughly 50\% of the total observation length.}. The data in 3$-$79 keV are cleaned and filtered for background events using the \nustar~Data Analysis Software (NUSTARDAS) version 1.4.1. Calibration files from {\tt NUSTAR CALDB}, upadated on 2014 November 14, are used for instrument responses. All light curves and spectra are extracted for the two focal plane modules (FPMA and FPMB) using the tool {\tt nuproducts}. The source spectra are extracted from a 30$^{\prime\prime}$  circular region centered on S5 0836+71. A circular region of 70$^{\prime\prime}$ radius is selected on the same chip, free from contaminating sources, to extract background spectra. The source spectra are binned to have at least 20 counts per bin to perform spectral fitting. Further, light curves are generated by summing FPMA and FPMB count rates, subtracting background, and using 4 ksec binning.

\subsection{{\it Swift} Observations}\label{subsec:swift}
The {\it Swift} satellite has observed S5 0836+71 multiple times with all the three instruments: the Burst Alert Telescope (BAT; \citealt{2005SSRv..120..143B}, 15$-$150 keV), the X-ray Telescope (XRT; \citealt{2005SSRv..120..165B}, 0.3$-$10 keV) and the UltraViolet Optical Telescope (UVOT; \citealt{2005SSRv..120...95R}) which can observe in six filters, namely, V, B, U, UVW1, UVM2 and UVW2.

The XRT data are first processed with the XRTDAS software package (v.3.0.0) available within the HEASOFT package (6.16). Following standard procedures ({\tt xrtpipeline v.0.13.0}), event files are cleaned and calibrated with the calibration database updated on 2014 November 7. Standard grade selections of 0$-$12 in the photon counting mode are used. Energy spectra are extracted from the summed event files. Source region is selected as 55$^{\prime\prime}$ circle while background region is chosen as an annular ring with inner and outer radii of 110$^{\prime\prime}$ and 210$^{\prime\prime}$ respectively, centered at the position of S5 0836+71. Moreover, when the source counts rate exceeds 0.5 counts s$^{-1}$, to avoid pile up effect, the source region is chosen as an annulus with inner and outer radii of 5$^{\prime\prime}$ and 65$^{\prime\prime}$ respectively, while the background spectra are extracted from an annular region of inner and outer radii 130$^{\prime\prime}$ and 230$^{\prime\prime}$ respectively \citep[see e.g.,][]{2013ApJS..207...28S}. The tool {\tt ximage} is used to combine exposure maps and the ancillary response files are generated using the task {\tt xrtmkarf}. Source spectra are binned to have at least 20 counts per bin, using the task {\tt grppha} and the spectral fitting is performed with XSPEC \citep{1996ASPC..101...17A}. An absorbed power law \citep[$N_{\rm H}$ = 2.83 $\times$ 10$^{20}$ cm$^{-2}$;][]{2005A&A...440..775K} is used for fitting and the uncertainties are calculated at 90\% confidence level.

{\it Swift}-UVOT observations are integrated using {\tt uvotimsum} and the magnitudes are extracted using the task {\tt uvotsource}. Source region is selected as a circle of 5$^{\prime\prime}$ radius centered at the source position, while the background is chosen from a nearby source-free circular region of 1$^{\prime}$ radius. Observed magnitudes are corrected for galactic extinction following \citet{2011ApJ...737..103S} and converted to flux units using the calibrations of \citet{2011AIPC.1358..373B}. These fluxes are also corrected for possible absorption by intervening Lyman-$\alpha$ absorption systems, following \citet{2010MNRAS.405..387G}.

\section{Results}\label{sec:results}
\subsection{Multi-band Temporal Variability}\label{subsec:mw_var}
\begin{figure*}
\hbox{
      \includegraphics[width=\columnwidth]{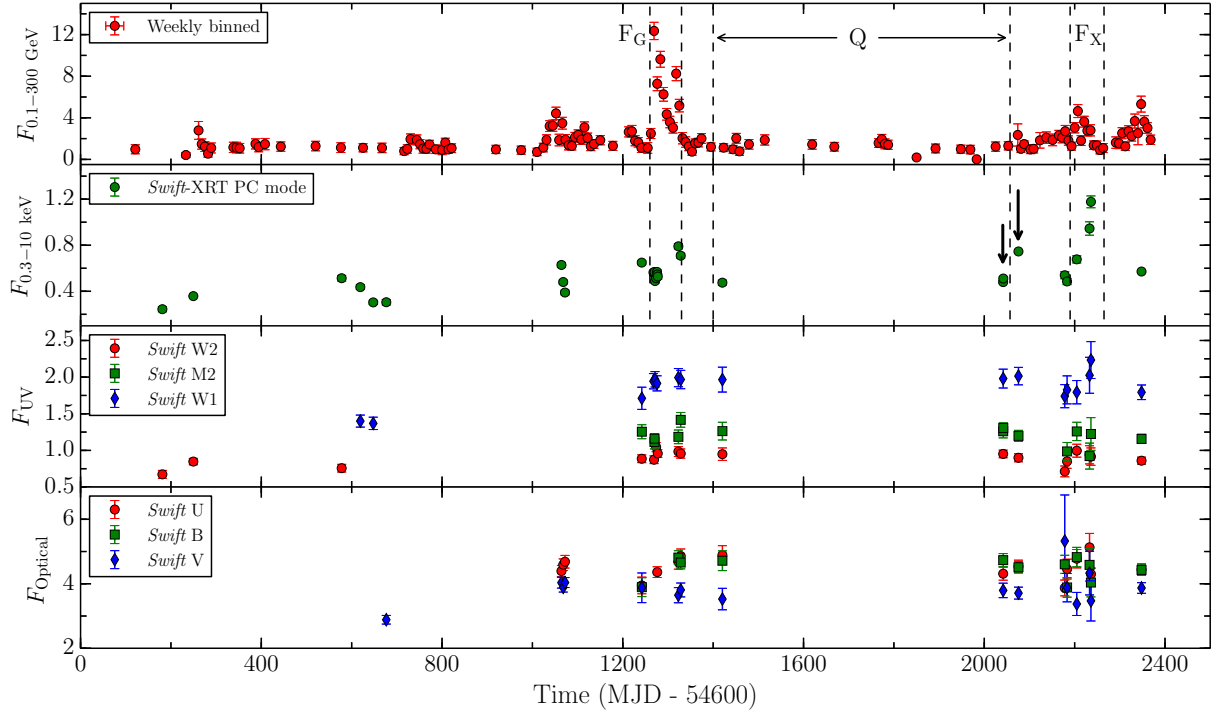}
     }
\caption{Multi-frequency light curves of S5 0836+71 since the launch of \fermi~satellite. \fermi-LAT and {\it Swift}-XRT data points are in units of 10$^{-7}$ \phflux~and counts s$^{-1}$ respectively. UV and optical fluxes have units of 10$^{-12}$ erg cm$^{-2}$ s$^{-1}$ . Periods corresponding to high $\gamma$-ray and X-ray states, and a low activity phase are quoted as F$_{\rm G}$, F$_{\rm X}$, and Q respectively. Black downward arrows represent the time of \nustar~observations. See text for details.}\label{fig:mw_lc}
\end{figure*}

The long term multi-band light curves of S5 0836+71, covering the period 2008 August 5 to 2014 November 15, are presented in Figure~\ref{fig:mw_lc}. In this plot, \fermi-LAT data points are weekly binned and {\it Swift} observations correspond to one point per observation ID. The flaring activities are seen both in X-ray and $\gamma$-ray bands. In comparison to high energy bands, available optical-UV observations show little or no flux variations. This suggests that even during the flaring periods, the optical-UV spectrum is accretion disk dominated, which is, in general, not expected to vary drastically. To substantiate this, fractional rms variability amplitude \citep[$F_{\rm var}$;][]{2003MNRAS.345.1271V} is calculated and the results are given in Table~\ref{tab:f_var}. This parameter is found to be largest for $\gamma$-rays and decreases with frequency, a trend generally found among blazars \citep[e.g.,][]{2005ApJ...629..686Z,2010ApJ...712..405V}.

It can be seen in Figure~\ref{fig:mw_lc} that the largest $\gamma$-ray flare, during the period MJD 55,860$-$55,930, is not associated with that in X-rays. The  enhancement in X-ray flux is relatively moderate at the time of the $\gamma$-ray flare, however, when X-ray flux peaks (MJD 56,790$-$56,865) there is mild increase in the $\gamma$-rays. To study this peculiar behavior, a separate $\gamma$-ray flaring period (F$_{\rm G}$; MJD 55,860$-$55,930) and X-ray flaring period (F$_{\rm X}$; MJD 56,790$-$56,865) are selected for further analysis. For comparison, a low activity state (Q; MJD 56,000$-$56,657) is also chosen. These periods are shown with dashed lines in Figure~\ref{fig:mw_lc}. The $\gamma$-ray flaring period selected in this work is same as that taken by \citet{2013A&A...556A..71A}.

\begin{figure*}
\hbox{
      \includegraphics[width=\columnwidth]{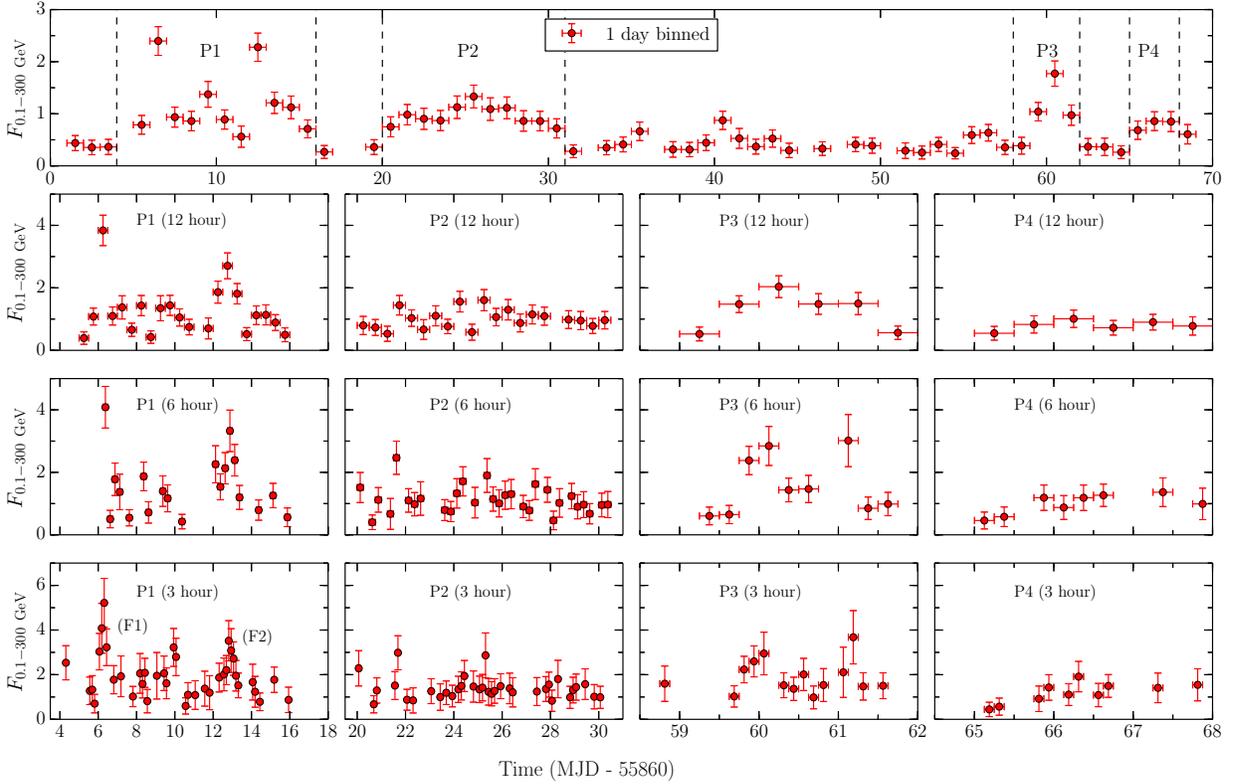}
     }
\caption{Gamma-ray light curves of S5 0836+71 covering the period of GeV outburst. Fluxes are in units of 10$^{-6}$ \phflux. Good photon statistics during this period allows to go for various time binnings. Top panel represents the daily binned light curve, from which, intervals of higher activities (marked as P1, P2, P3, and P4) are selected to go for twelve hours, six hours, and three hours binning. In the three hour binned light curve of P1 period (bottom left corner), two flares F1 and F2 are selected for profile fitting.}\label{fig:fermi_multi}
\end{figure*}

\begin{figure*}
\hspace{-1.0cm}
\hbox{
      \includegraphics[width=9cm]{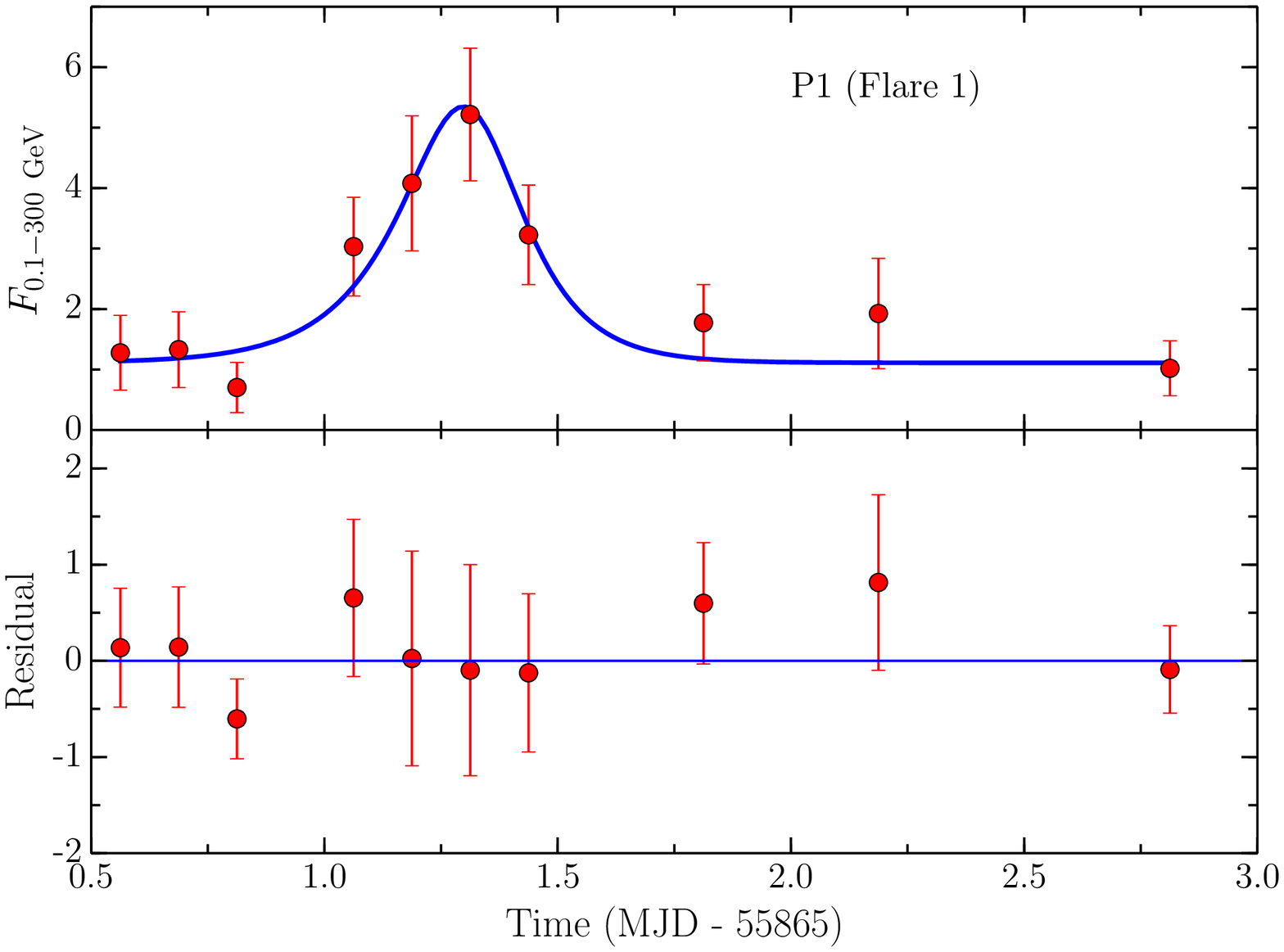}
      \includegraphics[width=9cm]{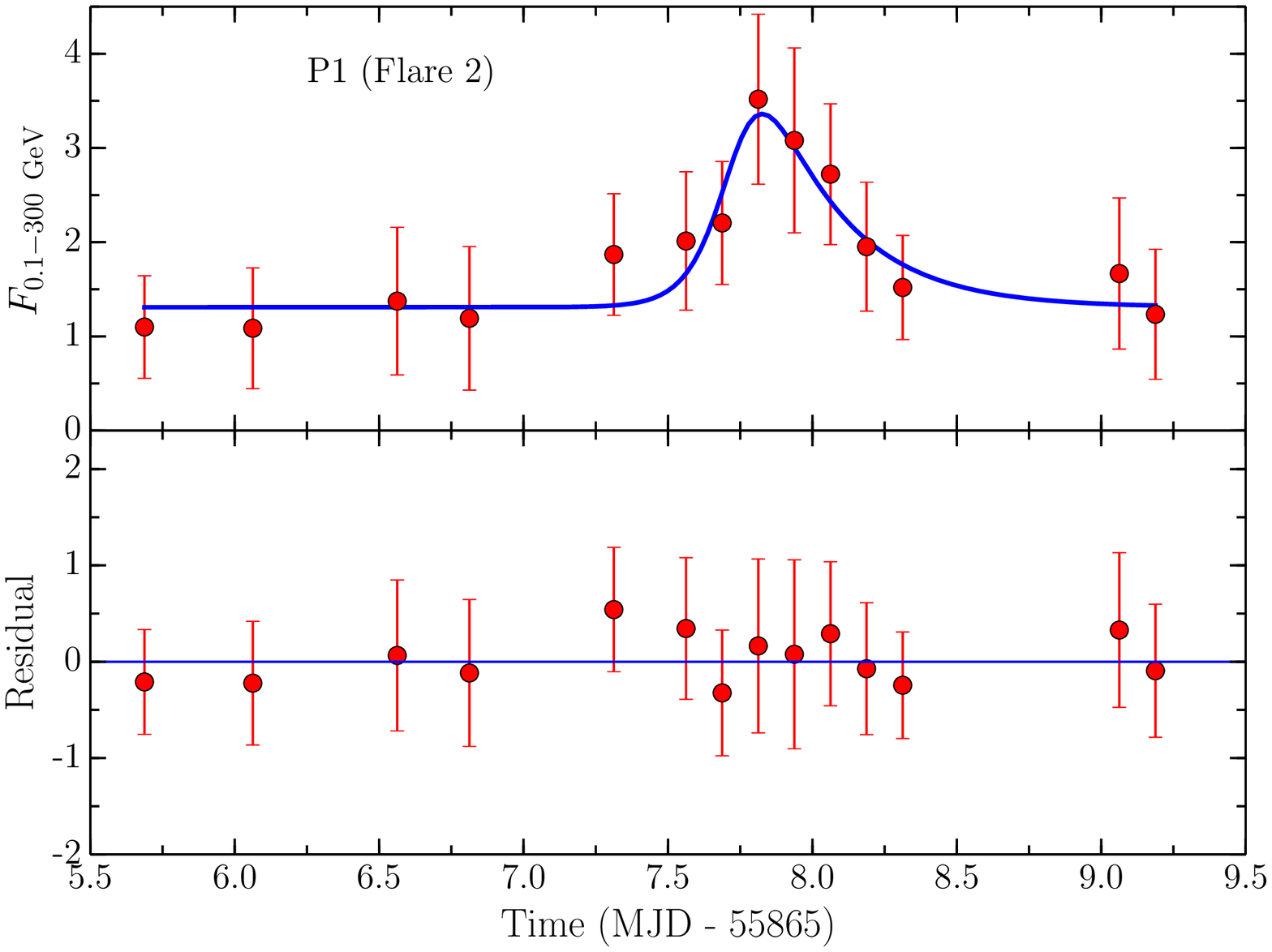}
     }
\caption{Top: Three hours binned $\gamma$-ray flares of S5 0836+71 selected for the time profile fitting, during the period of GeV outburst. Fluxes have same units as in Figure~\ref{fig:fermi_multi}. Blue solid line denotes the best-fit temporal profile assuming an exponential rise and fall. Bottom: The residual of the fitting.}\label{fig:flare_fit}
\end{figure*}

The good photon statistics during the period F$_{\rm G}$ allows to generate the one day binned $\gamma$-ray light curve, and is shown in the top panel of Figure~\ref{fig:fermi_multi}. In order to determine the shortest flux variability timescale and the highest $\gamma$-ray flux, the shorter duration flares are further selected from the one day binned $\gamma$-ray light curve to generate finer time binned light curves. They are annotated by P1, P2, P3, and P4 in Figure~\ref{fig:fermi_multi}. As can be seen, the largest flux enhancement is observed during the period P1. In the energy range of 0.1$-$300 GeV, the highest three hours binned $\gamma$-ray flux is (5.22 $\pm$ 1.10) $\times$ 10$^{-6}$ \phflux~measured on MJD 55,866, and the associated photon index is having a value of 2.62 $\pm$ 0.27. This corresponds to an apparent $\gamma$-ray luminosity of (1.62 $\pm$ 0.44) $\times$ 10$^{50}$ \lum. Within error, this luminosity is equal to the highest ever measured from a blazar \citep[2.1 $\pm$ 0.2 $\times$ 10$^{50}$ \lum~for 3C 454.3;][]{2011ApJ...733L..26A}. Further, to have a better estimate of the fastest rising and decaying timescales, two rapid flares are selected for time profile fitting. They are quoted as F1 and F2 in the period P1 in Figure~\ref{fig:fermi_multi} and are fitted using a rise and decay function given by \citep[see e.g.,][]{2010ApJ...722..520A}
\begin{equation}
F(t) = F_{\rm c} + F_0 \left[{\rm exp}\left(\frac{t_{\rm p} - t}{T_{\rm r}}\right) + {\rm exp}\left(\frac{t-t_{\rm p}}{T_{\rm d}}\right)\right]^{-1}
\end{equation}
where $F_{\rm c}$ decides the quiescent level and $F_0$ the amplitude of the flares, $t_{\rm p}$ measures the approximate time of the peak which is fixed at the time when the highest flux is observed, and $T_{\rm r}$ and $T_{\rm d}$ measure the rise and decay times. The results of the fitting are shown in Figure~\ref{fig:flare_fit} and the obtained parameters are provided in Table~\ref{tab:flare_fit}. The fastest flux rise and decay times are found to be $\sim$3 hours and $\sim$2.5 hours respectively. This is the first report of hour scale $\gamma$-ray variability detected from this source. Moreover, such short timescale flux variability was also probably not known in any high redshift blazar beyond redshift 2, prior to this work.

The maximum X-ray flux from S5 0836+71 is measured during the period F$_{\rm X}$. In the energy range of 0.3$-$10 keV and corrected for galactic absorption, it is having a value of 4.78$^{+0.57}_{-0.38}$ $\times$ 10$^{-11}$ \ergflux, detected on MJD 56,836. The associated photon index is hard and has a value of 1.47$^{+0.11}_{-0.11}$. Accordingly, the isotropic X-ray luminosity measured is $L_{\rm X}$ = 9.43 $\times$ 10$^{47}$ \lum.

\begin{figure*}
\hbox{
      \includegraphics[width=\columnwidth]{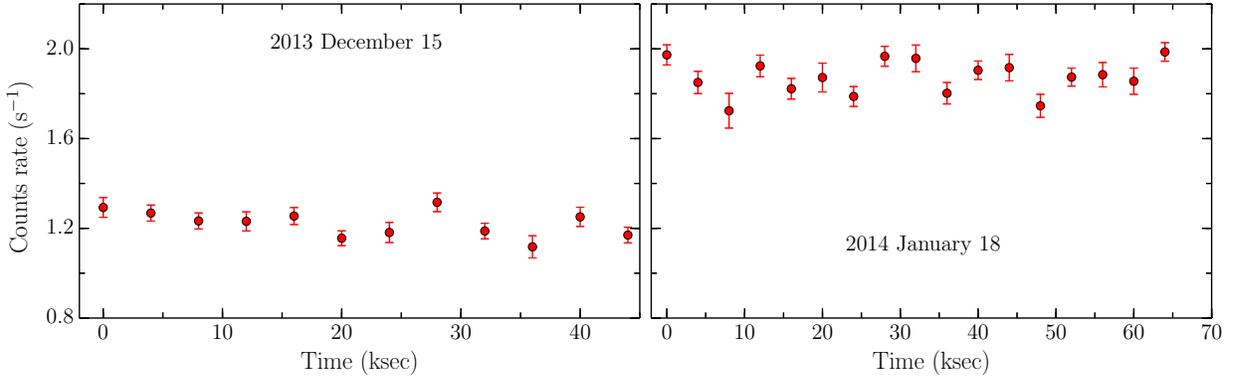}
     }
\caption{Background subtracted 3$-$79 keV light curves of the two \nustar~observations of S5 0836+71. For this plot, the FPMA and FPMB count rates are summed, and 4 ksec binning is applied.}\label{fig:nustar_lc}
\end{figure*}

The hard X-ray light curves, using the \nustar~observations, are generated with 4 ksec binning and shown in Figure~\ref{fig:nustar_lc}. The flux variations can be seen in both the light curves with the variability probability \citep{2010ApJ...722..520A} of 98.33\% and 99.86\% for 2013 December and 2014 January observations respectively. Moreover, there is significant change in the flux between the two epochs. Earlier, the hard X-ray flux of this source was found to vary on time scales of months using the {\it Swift}-BAT data \citep{2007ApJ...669..884S}. Better sensitivity of \nustar~has now enabled the detection of a shorter time scale of variability from S5 0836+71.

\subsection{Spectral Analysis}\label{subsec:spec_analysis}

The search for the presence of curvature in the $\gamma$-ray spectra is done in all the three activity states considered in this work. In particular, two spectral models are used: power law ($dN/dE \propto E^{\Gamma_{\gamma}}$), where $\Gamma_{\gamma}$ is the photon index and logParabola ({ $dN/dE \propto (E/E_{\rm o})^{-\alpha-\beta log({\it E/E_{\rm o}})}$}, where $E_{\rm o}$ is an arbitrary reference energy fixed at 300 MeV, $\alpha$ is the photon index at $E_{\rm o}$ and $\beta$ is the curvature index which defines the curvature around the peak). The test statistic of the curvature $TS_{\rm curve}$ = 2(log $\mathcal{L}$(LogParabola) $-$ log $\mathcal{L}$(power-law)), is calculated to test for the presence of curvature. Following \citet{2012ApJS..199...31N}, a threshold $TS_{\rm curve} >$ 16 is set to test the existence of a significant curvature. The resultant fitting parameters are given in Table~\ref{tab:gamma_spec}. Significant curvature is noticed only during the $\gamma$-ray flare with $TS_{\rm curve}$ = 51. Though at low significance, there is a hint for the presence of curvature in the X-ray flaring period also ($TS_{\rm curve} \approx$ 9), and during the low activity state the power law model represents the spectrum better.

\begin{figure*}
\hspace{-1.0cm}
\hbox{
      \includegraphics[width=9cm]{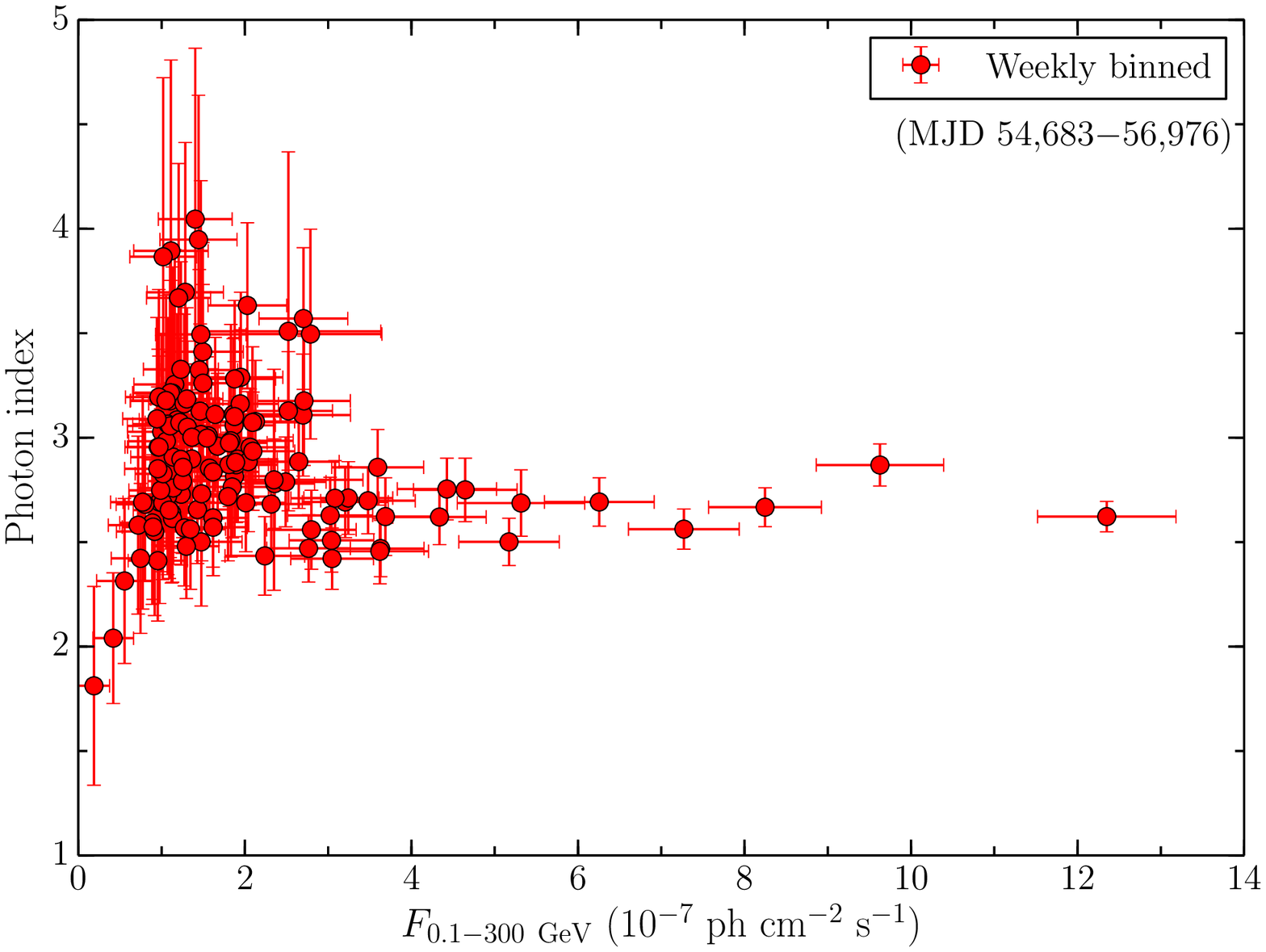}
      \includegraphics[width=9cm]{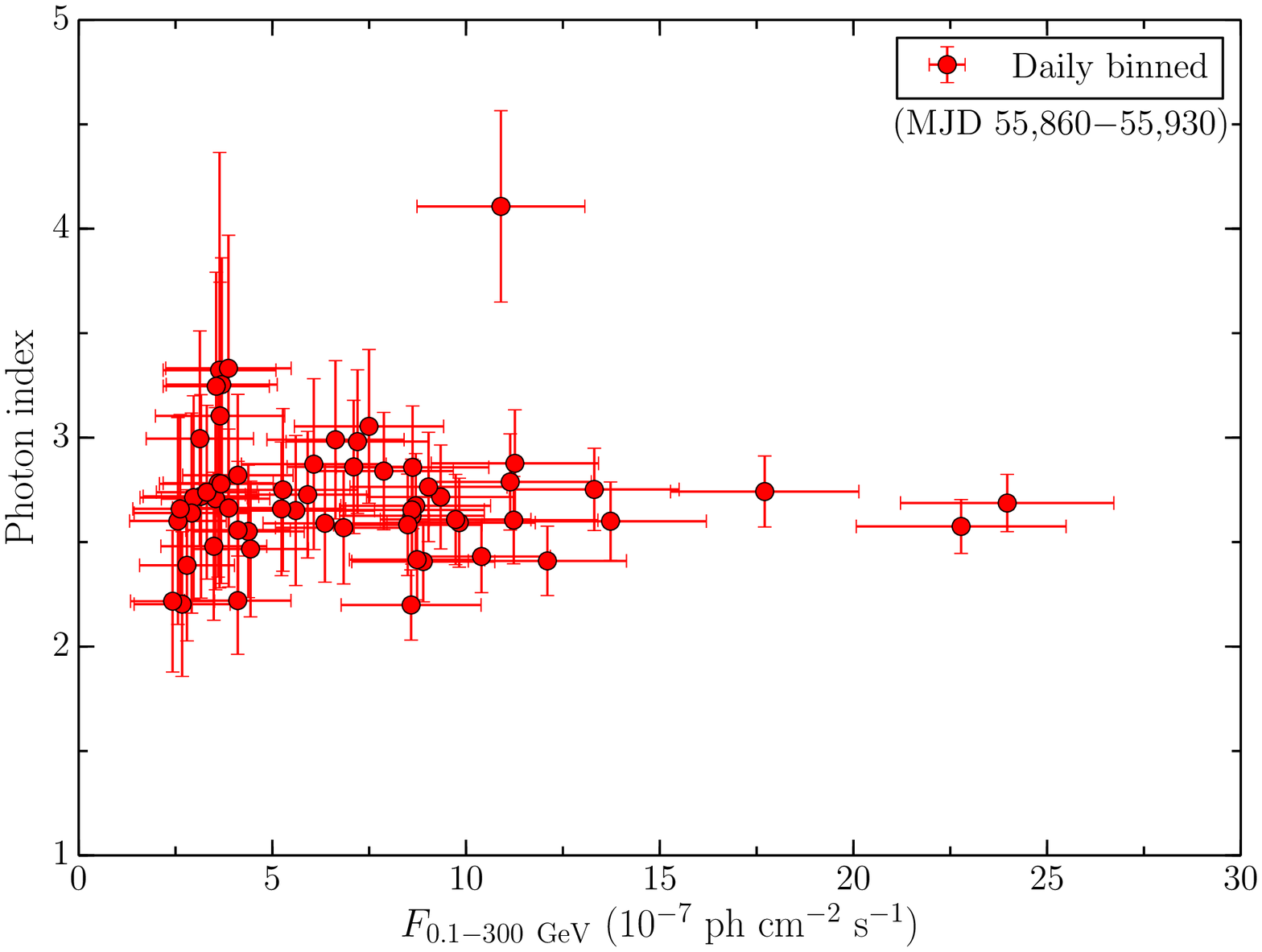}
     }
\caption{Left: Weekly scatter plot of the $\gamma$-ray flux vs. photon index. Right: Same as left but for daily binned data covering the period of high $\gamma$-ray activity.}\label{fig:flux_index}

\end{figure*}

To study the overall $\gamma$-ray spectral behavior, the photon index is plotted against the flux in Figure~\ref{fig:flux_index}. In this figure, the variation of photon indices with fluxes is shown both for the weekly binned $\gamma$-ray light curve of the entire six years (Figure~\ref{fig:mw_lc}) and the daily binned $\gamma$-ray light curve covering the period of GeV outburst (Figure~\ref{fig:fermi_multi}). Visual inspection of the daily binned data does not reveal any correlation, however, for weekly binned plot the source seems to show a `softer when brighter' trend, up to a flux level of $\simeq$1.5 $\times$ 10$^{-7}$ \phflux. Above this flux value, there is a hint for a `harder when brighter' behavior. To statistically confirm these findings, a Monte Carlo simulation is performed that takes into account of the dispersion in the flux and photon index measurements. In particular, for each observed randomly drawn pair of flux and photon index value, the data is re-sampled by extracting it from a normal distribution centered on the observed value and having standard deviation equal to the 1$\sigma$ error estimate. For daily binned data set, the correlation coefficient ($\rho$) is 0.02 $\pm$ 0.13 with a 95\% confidence limit -0.23 $\leqslant$ $\rho$ $\leqslant$ 0.28, clearly suggesting the absence of any correlation between fluxes and photon indices. For lower fluxes of weekly binned data (i.e., $F_{\gamma}$ $<$ 1.5 $\times$ 10$^{-7}$ \phflux), $\rho$ is found to be 0.18 $\pm$ 0.12 with 95\% confidence limit of -0.07 $\leqslant$ $\rho$ $\leqslant$ 0.41. For higher flux level ($F_{\gamma}$ $>$ 1.5 $\times$ 10$^{-7}$ \phflux), $\rho$ is found to be  0.27 $\pm$ 0.11 with 95\% confidence limit of 0.02 $\leqslant$ $\rho$ $\leqslant$ 0.48. Thus, based on the Monte Carlo analysis, claim for the presence of correlation between fluxes and photon indices in both the weekly binned and daily binned data set, as shown in Figure~\ref{fig:flux_index}, cannot be made.

\begin{figure*}
\hspace{-1.0cm}
\hbox{
      \includegraphics[width=9cm]{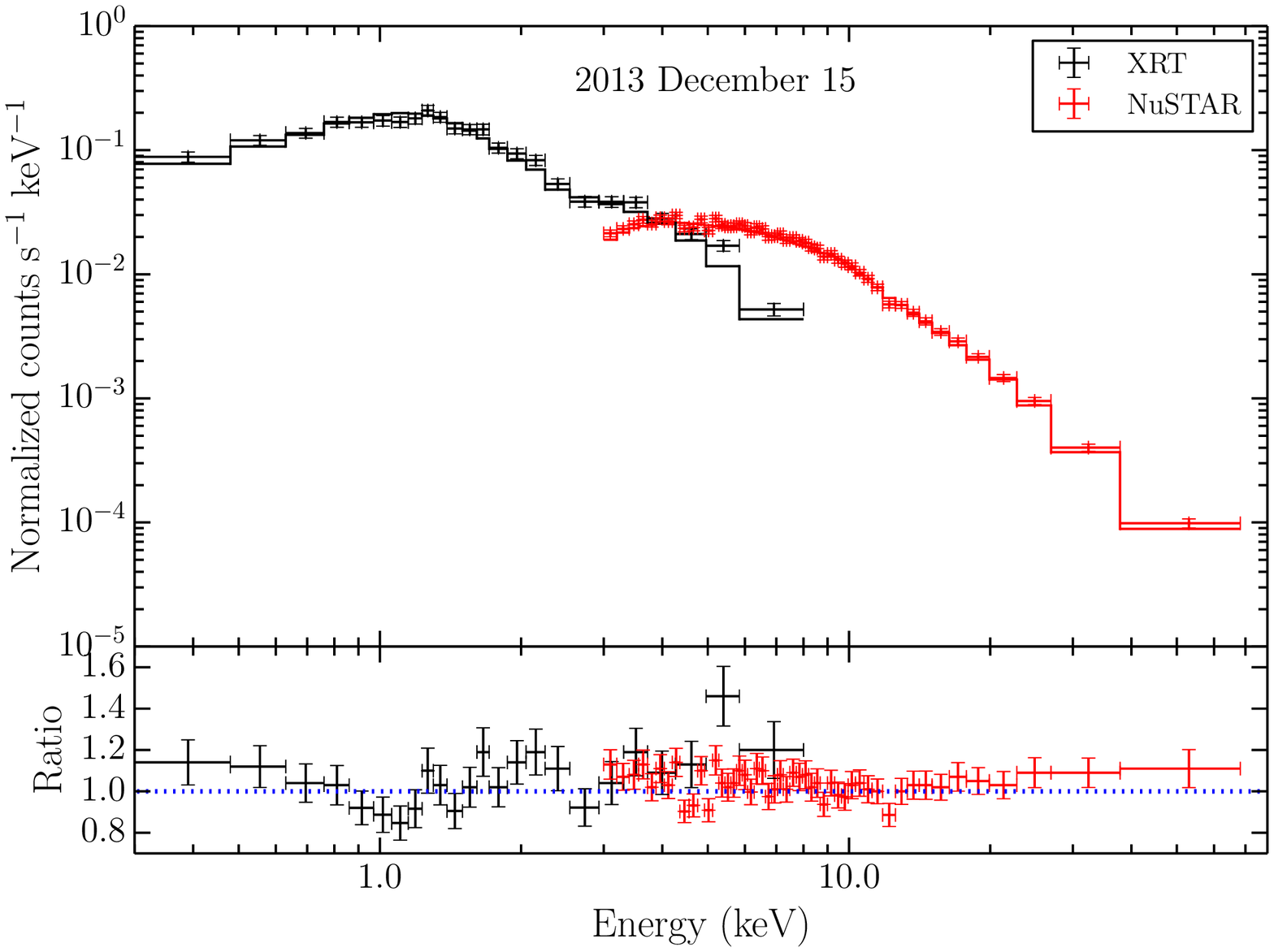}
      \includegraphics[width=9cm]{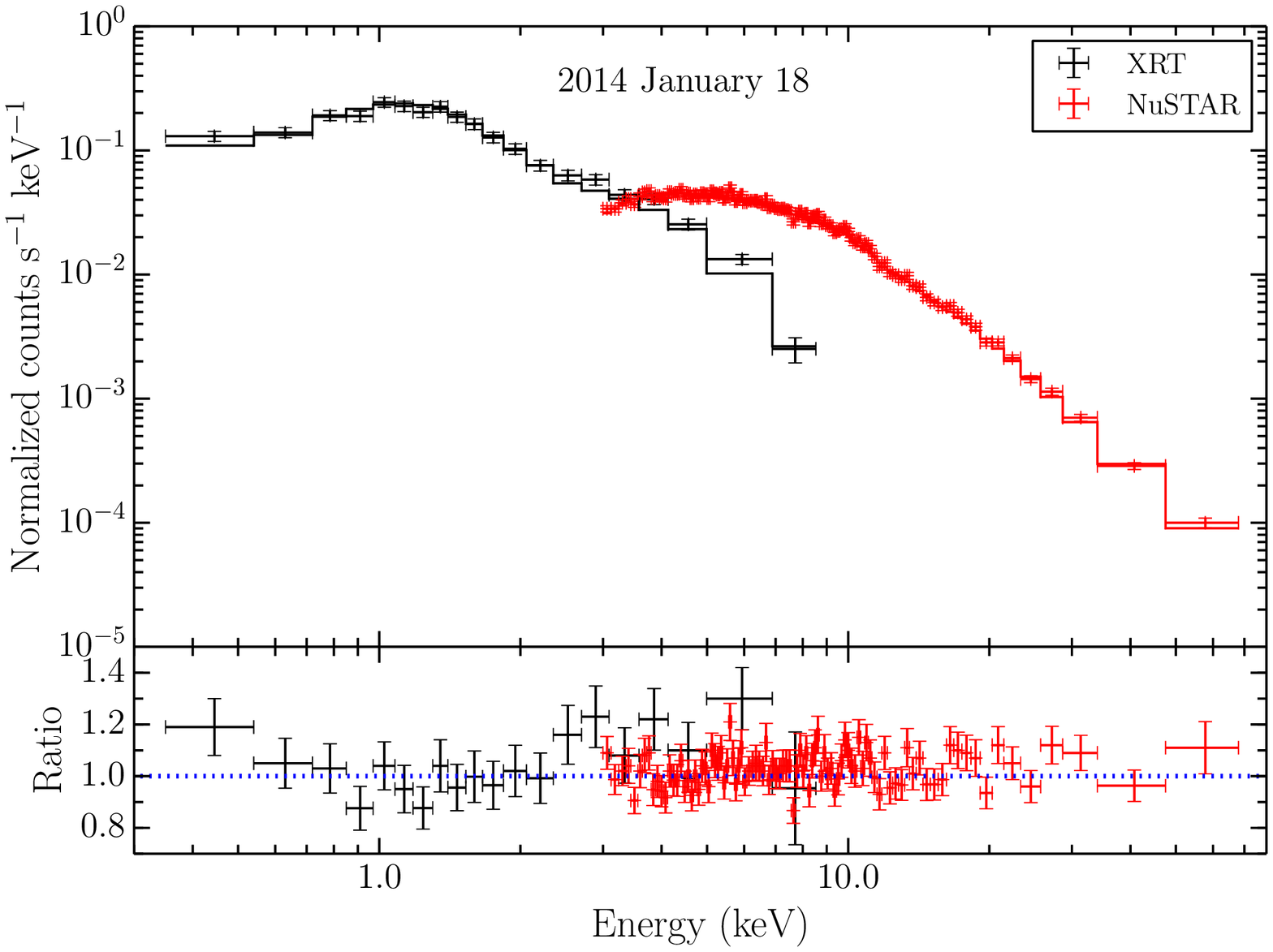}
     }
\caption{Top: Joint XRT (0.3$-$10 keV) and \nustar~(3$-$78 keV) spectrum, fit with power law plus two layers of absorption. Bottom: data to model ratio for the obtained fit. Two \nustar~spectra are grouped in XSPEC and rebinned for visual clarity.}\label{fig:xrt_nustar}

\end{figure*}

Since {\it Swift} has monitored S5 0836+71 simultaneously with \nustar~in 2013 December and 2014 January (shown by black downward arrows in Figure~\ref{fig:mw_lc}), joint spectral fit to the XRT and \nustar~data is also attempted. The difference in flux calibration between XRT and \nustar~spectra is allowed to vary by including an intercalibration constant (CONST in XSPEC). This constant is fixed to 1 for two \nustar~spectra, since the calibration difference between FPMA and FPMB are on the order of 1\% \citep[see e.g.,][]{2014ApJ...787...83M}, and left free to vary for the XRT. A first round of fitting is performed with a power law model plus fixed galactic absorption which clearly does not provide acceptable fitting parameters (see Table~\ref{tab:xrt_nustar}; the data/model ratio is shown in Figure~\ref{fig:ratio_abspl}). However, adding intrinsic absorption ({\tt zwabs} in XSPEC) or a broken power law shape for the continuum, improves the fit (F-test probability of null hypothesis $<$ 10$^{-10}$). For the case of intrinsic absorption, the level of absorption is higher but with larger uncertainty, and the obtained $N_{\rm H}$ value is similar to that reported by \citet{2007ApJ...669..884S}. Testing the broken power law model on joint XRT-\nustar~spectra, while keeping the Galactic absorption fixed, also results in a satisfactory fit. Moreover, following \citet{2013A&A...556A..71A}, fitting the data with a power law plus two layers of galactic absorption (keeping one absorption fixed to the Galactic value and other left to vary) also gives a good fit with statistical parameters similar to that obtained with the use of broken power law model (see Table~\ref{tab:xrt_nustar}). In this case, the additional absorption is higher than the Galactic absorption by a factor of $\sim$3. Comparison of the fitting obtained with broken power law model with that obtained from power law plus two layers of absorption, are inconclusive. The F-test probability of null hypothesis, that the data is better described by the power law plus two layers of absorption, is found to be 0.014 and 0.001 for 2013 December and 2014 January observations respectively. In Figure~\ref{fig:xrt_nustar}, the results of fitting the power law with two layers of absorption for two simultaneous XRT-\nustar~observations are shown.

\subsection{Spectral Energy Distributions}\label{subsec:sed}
\subsubsection{Model Setup}\label{subsubsec:model}
To reproduce the broadband SEDs of S5 0836+71, a simple one zone leptonic emission model is developed, by following the prescriptions of \citet{2009MNRAS.397..985G} and \citet{2009ApJ...692...32D} \citep[see also][]{2008ApJ...686..181F}, and is briefly described here. The emission region is assumed to be located at a distance of $R_{\rm diss}$ from the central black hole, and filled with electrons having smooth broken power law energy distribution
\begin{equation}
 N'(\gamma')  \, = \, N'_0\, { (\gamma'_{\rm b})^{-p} \over
(\gamma'/\gamma'_{\rm b})^{p} + (\gamma'/\gamma'_{\rm b})^{q}},
\end{equation}
 where $p$ and $q$ are the particle indices before and after the break energy ($\gamma'_{\rm b}$) respectively (primed quantities are measured in the comoving frame). The emission region is assumed to be spherical and its size is constrained by considering it to cover the entire jet cross-section with jet semi opening angle of 0.1 radians. A standard accretion disk is assumed \citep{1973A&A....24..337S} and it produces a multi-temperature blackbody spectrum \citep{2002apa..book.....F}. Above and below the accretion disk, the presence of X-ray corona is also considered which reprocesses 30\% of the accretion disk luminosity. Its spectrum is considered to be a cut-off power law: $L_{\rm cor}(\epsilon)\propto \epsilon^{-\alpha_{\rm cor}}\exp(-\epsilon/\epsilon_{\rm c})$. The BLR is assumed to be a spherical shell located at a distance of $R_{\rm BLR} = 10^{17} L^{1/2}_{\rm d,45}$ cm, where $L_{\rm d,45}$ is the accretion disk luminosity in units of 10$^{45}$ \lum. It reprocesses 10\% of the accretion disk luminosity. The SED of the BLR is approximated as an isotropic blackbody peaking at the rest-frame frequency of the Lyman-$\alpha$ line \citep{2008MNRAS.386..945T}. The dusty torus is assumed to be a spherical shell located at a distance $R_{\rm torus} = 10^{18} L^{1/2}_{\rm d,45}$ cm, reprocessing 50\% of the accretion disk radiation in the infrared. The spectrum of the torus is assumed as a blackbody with temperature $T_{\rm torus}$ = $\epsilon_{\rm peak}m_{\rm e}c^2/3.93k$, where $\epsilon_{\rm peak}$ is the dimensionless peak photon energy and $k$ is the Boltzmann constant. The relative contributions of these emissions with respect to the distance from the central black hole are calculated following the methodology of \citet{2009MNRAS.397..985G}. In the presence of uniform but tangled magnetic field, electrons radiate via synchrotron and SSC mechanisms \citep{2008ApJ...686..181F}. The external Compton emissions are also calculated with seed photons coming from the accretion disk, the BLR, and the dusty torus \citep{2009MNRAS.397..985G,2009ApJ...692...32D,2009herb.book.....D}. The kinetic power of the jet is computed by assuming protons to be cold, thus contributing only to the inertia of the jet, and having equal number density to that of the relativistic electrons \citep[e.g.,][]{2008MNRAS.385..283C}.

\subsubsection{SED Modeling Results}\label{subsubsec:sed_results}
\begin{figure*}
\hbox{\hspace{-1.5cm}
      \includegraphics[width=6.5cm]{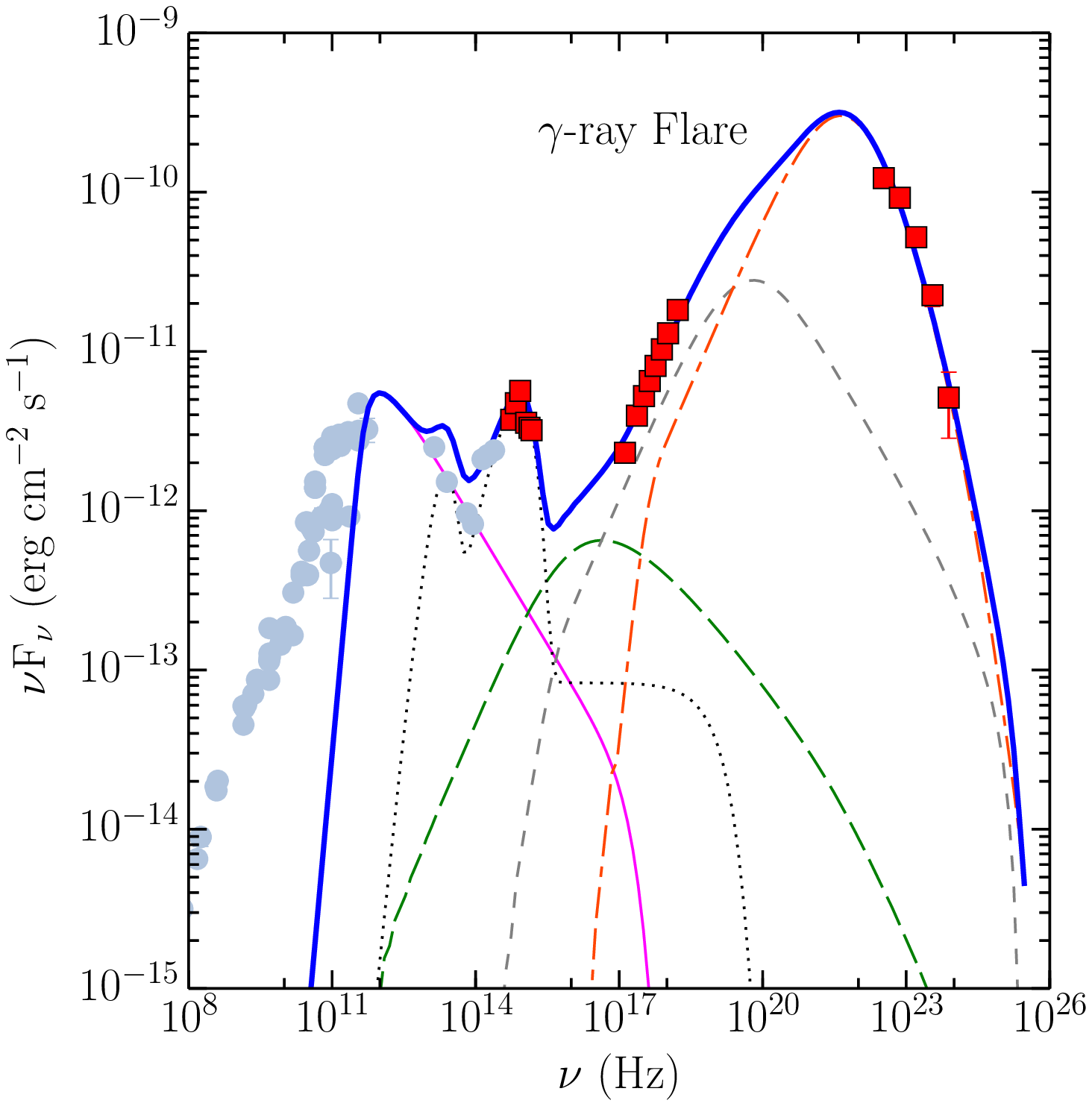}
      \includegraphics[width=6.5cm]{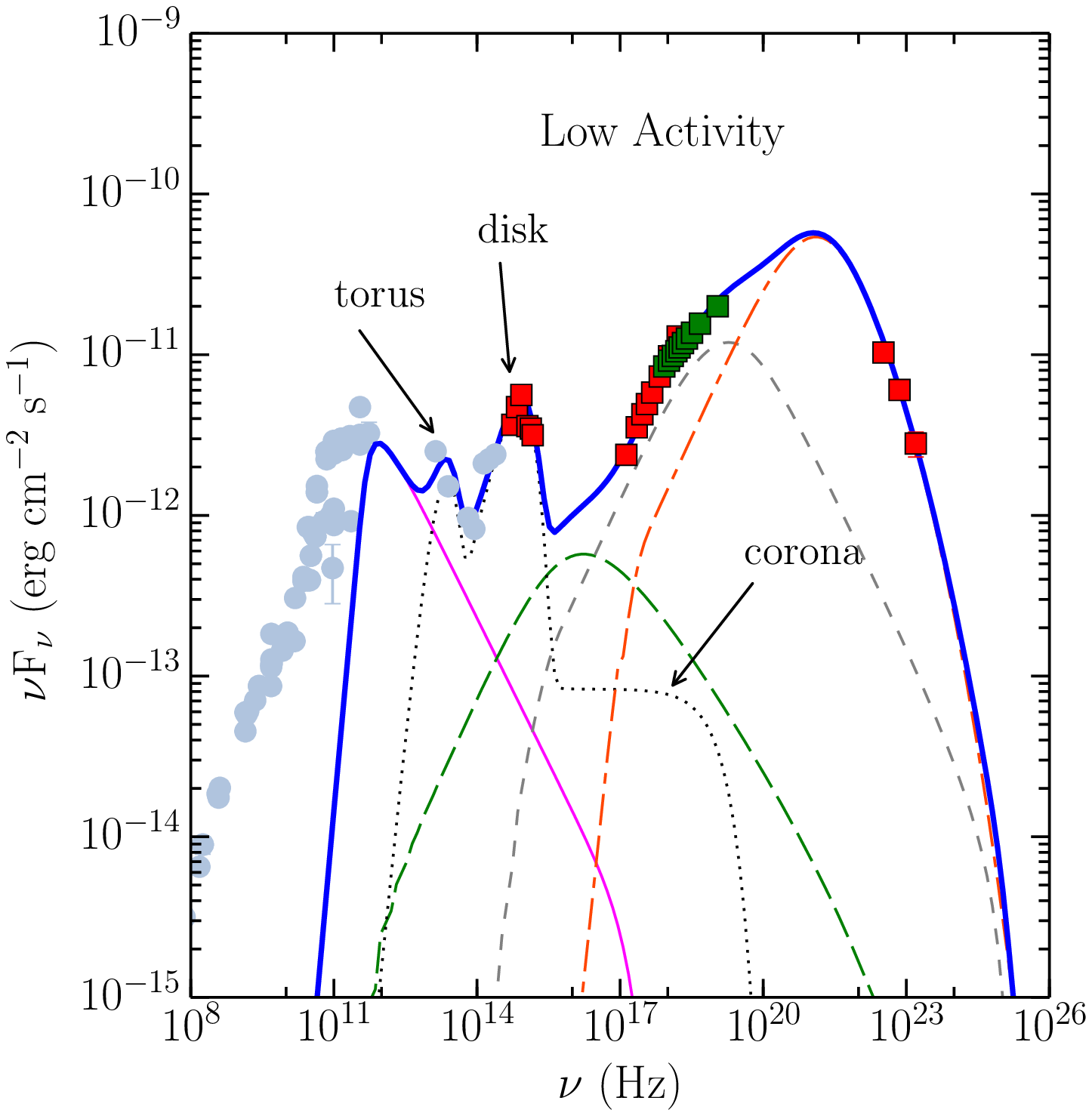}
      \includegraphics[width=6.5cm]{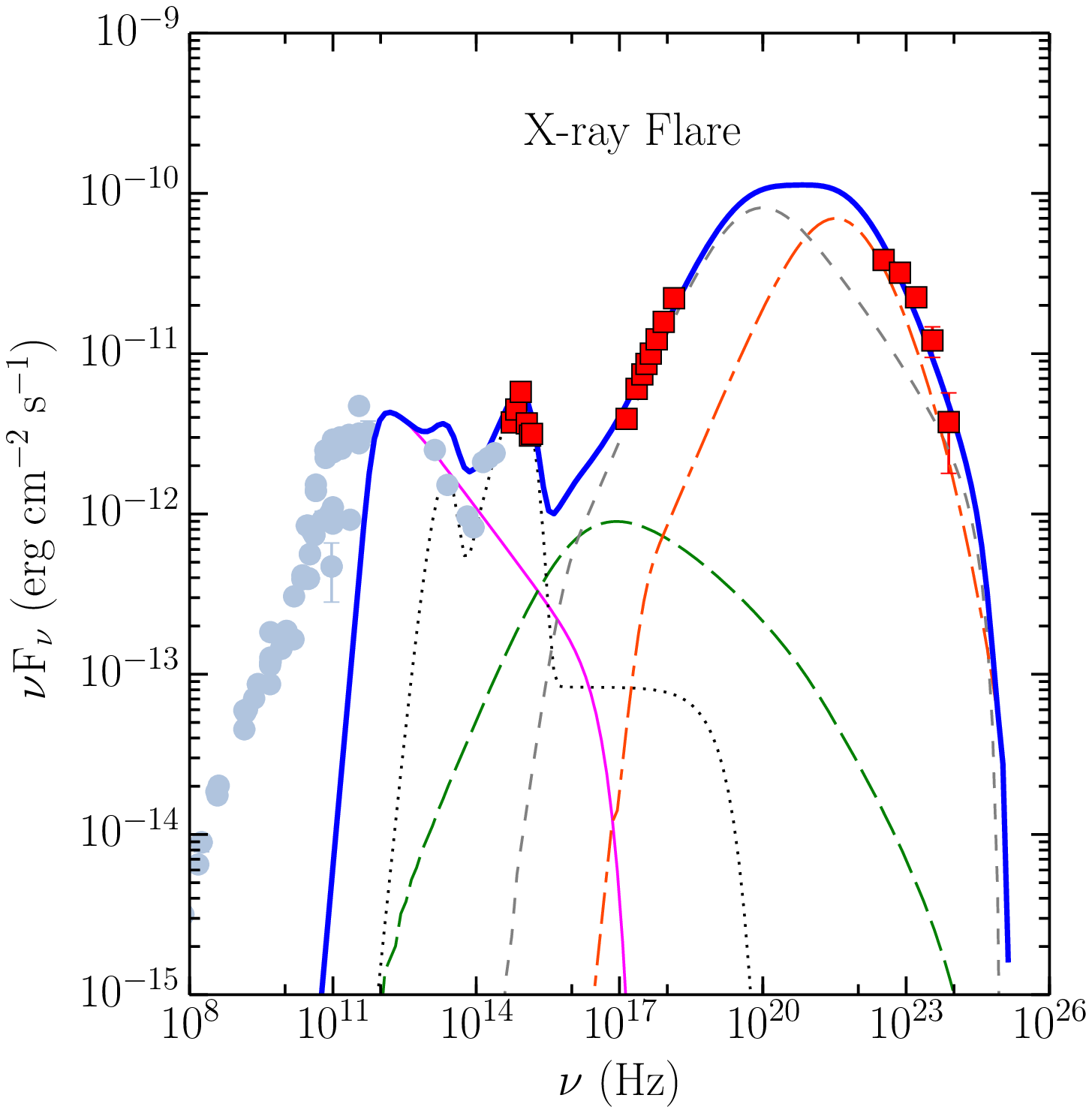}
     }
\hbox{\hspace{-1.5cm}
      \includegraphics[width=6.5cm]{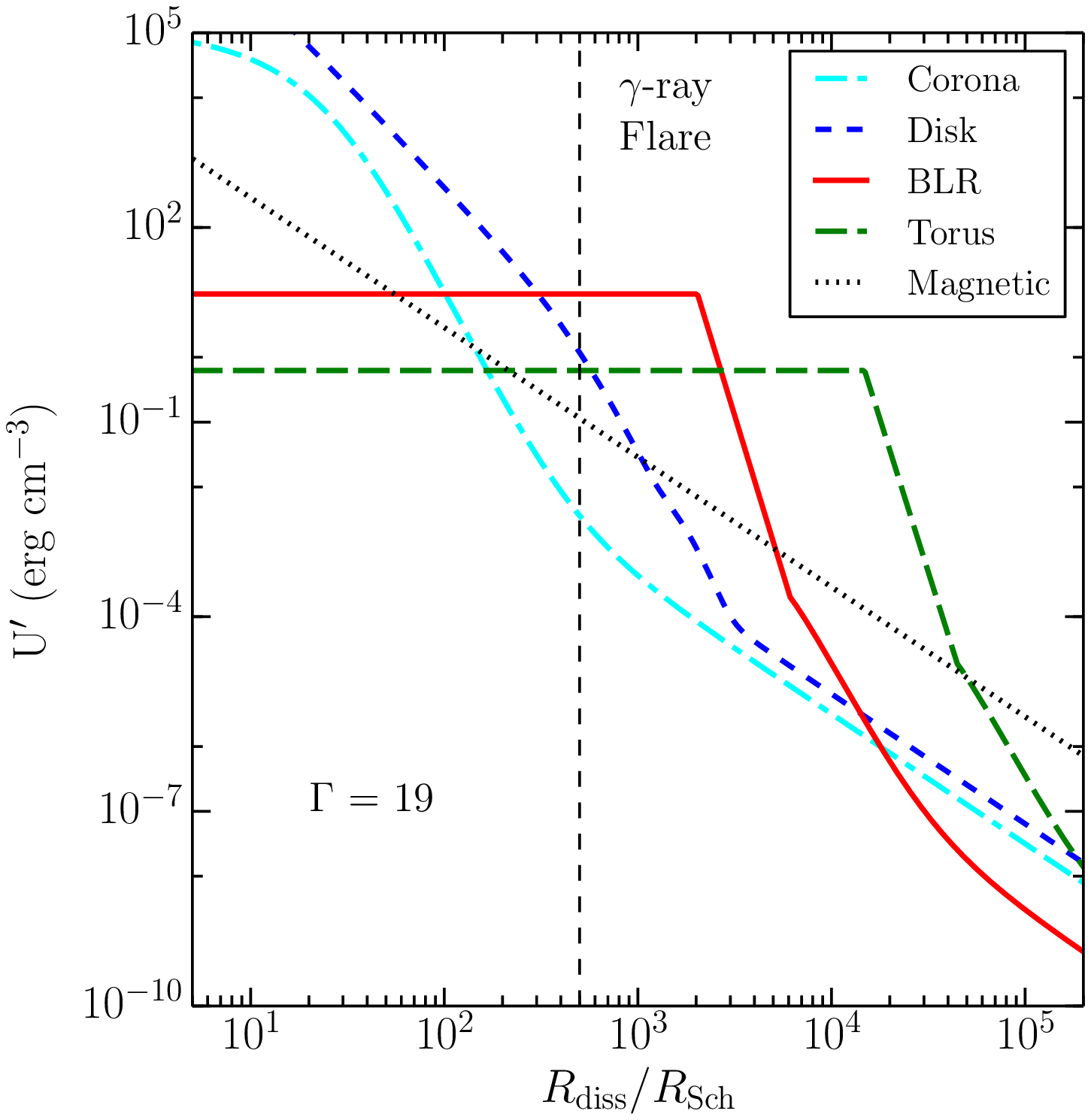}
      \includegraphics[width=6.5cm]{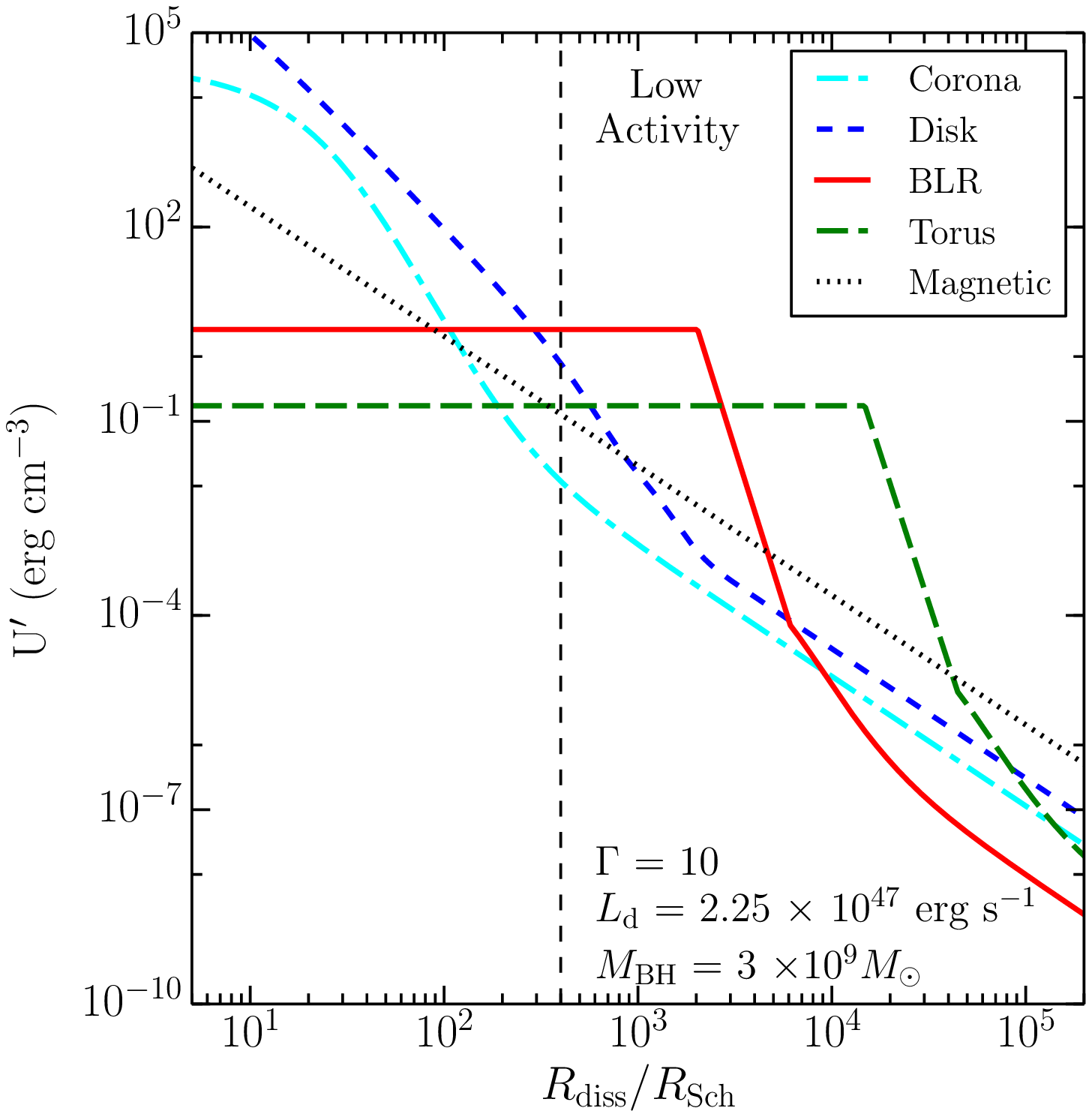}
      \includegraphics[width=6.5cm]{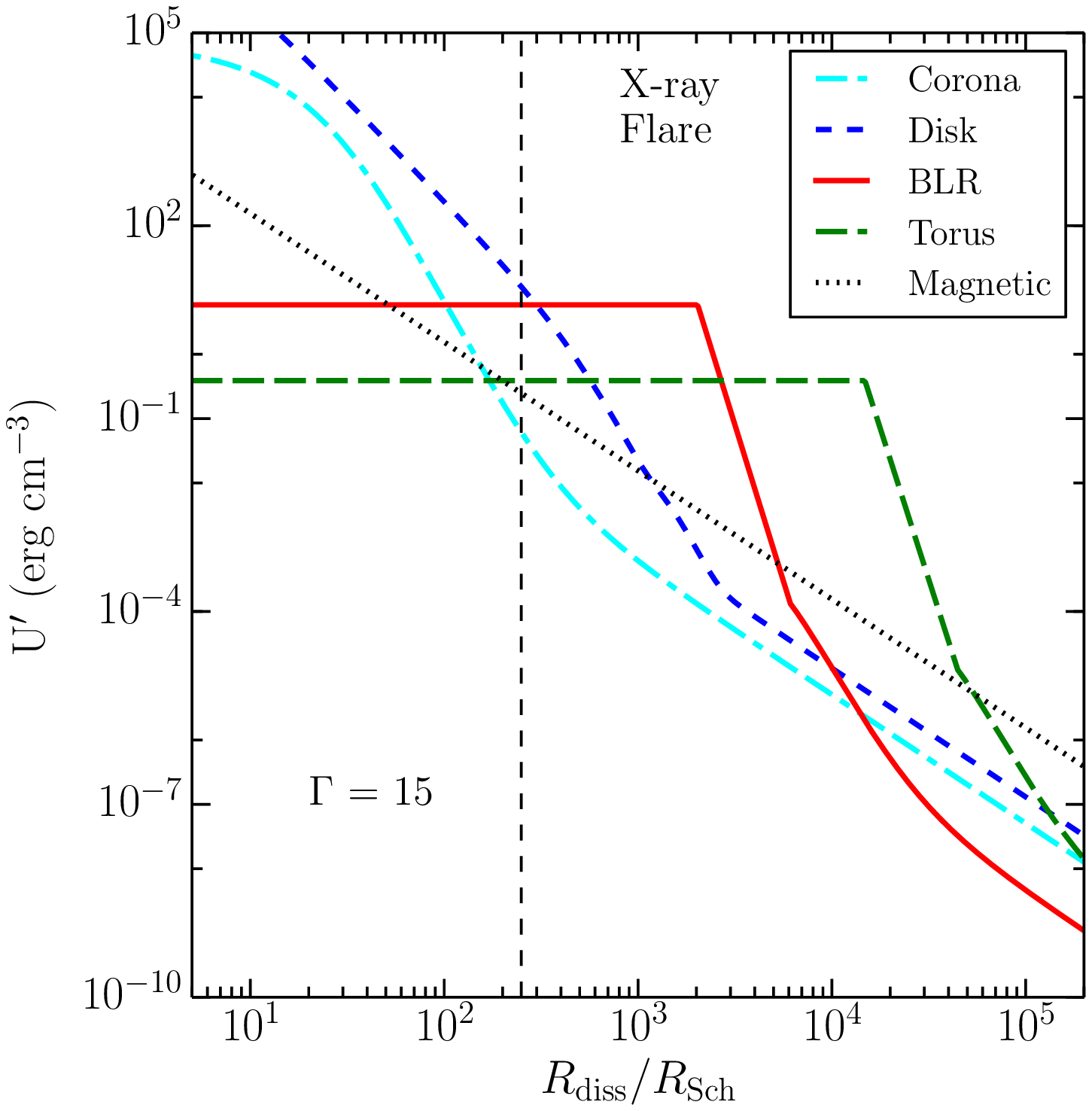}
     }
\caption{Top: Spectral energy distributions of S5 0836+71 during different activity states. Simultaneous {\it Swift} and \fermi-LAT data are shown with red squares and light blue circles represent archival observations. During the low activity period, the \nustar~data is also available which is shown by green squares. Pink thin solid and green long dashed lines correspond to synchrotron and SSC respectively. EC-disk and EC-BLR mechanisms are shown by grey dashed and orange dash-dash-dot lines respectively. Blue thick solid line is the sum of all the radiative contributions. Bottom: Variations of the energy densities measured in the comoving frame, as a function of the distance from the black hole, in units of $R_{\rm Sch}$. Vertical line denotes the location of the emission region.}\label{fig:sed_model}
\end{figure*}

The SED of S5 0836+71 is generated during three different activity states. The fluxes are averaged for the time intervals shown in Figure~\ref{fig:mw_lc} and the derived values are given in Table~\ref{tab:sed_flux}. The broadband SEDs are generated and modeled using the guidelines presented in Section~\ref{subsubsec:model}. The results are shown in Figure~\ref{fig:sed_model} and the associated parameters are presented in Table~\ref{tab:sed_par}. Variations of the energy densities, calculated in the comoving frame, are also shown in the bottom panels of Figure~\ref{fig:sed_model}.

Two crucial parameters for SED modeling of FSRQs, the accretion disk luminosity and the black hole mass, can be precisely constrained by fitting a standard accretion disk model \citep{1973A&A....24..337S} to the optical-UV spectrum, provided the big blue bump is visible \citep[e.g.,][]{2010MNRAS.405..387G}. Though this feature is clearly visible in all the activity states of S5 0836+71, the accretion disk model is used to reproduce the low activity optical-UV spectrum, as during the flaring period there could be a possibility of contamination from non-thermal jetted emission. The obtained disk luminosity and black hole mass are 2.25 $\times$ 10$^{47}$ \lum~and 3 $\times$ 10$^{9}$ \Msun~respectively, which are in perfect agreement with that obtained by \citet{2010MNRAS.405..387G}. Accordingly, the accretion disk luminosity is $\sim$60\% of Eddington luminosity.

Although the observed $\gamma$-ray variability timescales are a fraction of day, longer integration times are required to produce good quality LAT SEDs for spectral modeling. The fluxes at other frequencies are also averaged during the $\gamma$-ray integration period. Therefore, parameters obtained from the generation and modeling of the SEDs should be considered as representative of the average characteristics of S5 0836+71 during the various activity states.

\section{Discussion}\label{sec:dscsn}
In this work, major emphasis is given on the $\gamma$-ray temporal properties of S5 0836+71 during its GeV outburst in 2011 and on the physical properties of the source via SED modeling. The results of the monitoring from the sensitive hard X-ray telescope \nustar~are also presented for the first time.

The long term multi-wavelength study of S5 0836+71 is done in great detail by \citet{2013A&A...556A..71A}. They report a weak optical-$\gamma$-ray flux correlations. In this work also, though data is too sparse, visual inspection suggests the absence of any analogy between $\gamma$-ray and optical fluxes. This could be primarily due to the fact that optical-UV spectrum of S5 0836+71 is dominated by the accretion disk radiation, which is, in general, unlikely to vary on short timescales. An interesting feature of the multi-frequency light curve is the variation of X-rays as compared to $\gamma$-rays. Though the fluxes have enhanced during different flaring periods, the amplitude of variability does not follow the same pattern (Figure~\ref{fig:mw_lc}). At the time of GeV outburst, X-ray flux variations are relatively moderate, and when the X-ray flux is highest, there is relatively mild enhancement in $\gamma$-rays. This is explained by locating the emission region at different jet environments where various radiative energy densities play major role in describing the data, as discussed later in this section. 

The weekly binned $\gamma$-ray light curve, presented in the top panel of Figure~\ref{fig:mw_lc}, show multiple episodes of flaring activities with the brightest flare occurred during 2011. Apart from that, the source is sporadically detected throughout the \fermi~operation. The good photon statistics during the highest $\gamma$-ray activity period makes possible to study this event in detail. A daily binned $\gamma$-ray light curve is generated which shows the presence of sub-flares within the time period of enhanced emission (Figure~\ref{fig:fermi_multi}). From the one day binned light curve, further periods of high activity are selected (P1, P2, P3, and P4) and close-up light curves are generated with finer time bins. As can be seen, the flares have complex structure and asymmetric profiles and in few cases significant variations are visible. Results of the flare profile fitting indicate that the flare F1 seems to have a symmetric profile and the flare F2 possess asymmetric shape with fast rise and slow decay. Such a fast rise of the flare could be associated with the faster acceleration, probably at shock front, and the slow decay can be attributed to the weakening of the shock.

The location of the emission region, obtained from the SED modeling, is found to be inside the BLR where the energy density provided by the BLR clouds should dominate the energy densities of other photon fields in the jet rest frame, and hence the primary mechanism for the production of $\gamma$-rays would be IC upscattering of the BLR photons (observed energies $\epsilon_0 \simeq$ 10 eV). Accordingly, the cooling timescale for the electrons emitting $\gamma$-ray radiation with the energies $\epsilon_{\gamma} \simeq$ 1 GeV, measured in the observer's frame, would be
\begin{equation}
\tau_{\rm cool} \simeq \frac{3m_{e}c}{4\sigma_{\rm T}u'_{\rm BLR}} \sqrt{\frac{\epsilon_0(1+z)}{\epsilon_{\gamma}}},
\end{equation}
i.e., $\sim$9 minutes for the comoving BLR photon energy density $u'_{\rm BLR} \simeq$ 10 erg cm$^{-3}$. The calculated cooling time is significantly shorter than the observed shortest flux decay time of the flares, implying that the observed flux decrease is governed by processes other than radiative energy losses, probably a combination of different factors such as particle acceleration or jet dynamics \citep[e.g.,][]{2009ApJ...692.1374B,2014MNRAS.442..131K}. On the other hand, geometry and presence of sub-structures in the emitting region could also led to such observations \citep[][]{2001ApJ...563..569T}.

At the peak of the $\gamma$-ray flare, the highest measured isotropic $\gamma$-ray luminosity is $\sim$1.64 $\times$ 10$^{50}$ \lum, which corresponds to the total power emitted in the $\gamma$-ray band, in the proper frame of the jet, $L_{\gamma, \rm em} \simeq L_{\gamma}/2\Gamma^{2} \simeq$ 2.27 $\times$ 10$^{47}$ \lum, assuming the bulk Lorentz factor $\Gamma$ = 19 obtained from the modeling of the $\gamma$-ray flaring SED. This is a substantial fraction of the kinetic jet power ($\sim$15 \%, $P_{\rm j,kin}$ = 1.48 $\times$ 10$^{48}$ \lum.), implying the jet becomes radiatively efficient and the bulk of the radiative energy released in the form of $\gamma$-rays. For a comparison, the $L_{\gamma, \rm em}$ is found to be about 60\% of the Eddington luminosity. Moreover, $L_{\gamma, \rm em}$ is also a good fraction of the entire available accretion power ($L_{\rm acc} \simeq L_{\rm disk}/\eta_{\rm disk} = 2.25 \times 10^{48}$ erg s$^{-1}$; assuming radiative efficiency $\eta_{\rm disk}$ = 10\%).

The results on energetics suggest that both the total jet power (dominated by protons), and $L_{\rm acc}$ exceed the Eddington luminosity. However, there could be few possible caveats. For a given value of $L_{\rm disk}$, a relatively higher $\eta_{\rm disk}$ will reduce the total accretion power. Presence of more than one leptons per proton in the jet will decrease the budget of the proton jet power. Most importantly, underestimation of the central black hole mass can also lead to super Eddington accretion and jet power. The mass of the central black hole used in this work is 3 $\times$ 10$^9$ $M_{\odot}$, which is obtained by fitting the standard accretion disk model to the optical-UV spectrum. However, if the C~{\sc iv} line parameters from \citet{2012RMxAA..48....9T} are used to calculate the black hole mass of S5 0836+71 and the virial relations of \citet{2011ApJS..194...45S} are adopted, the estimated black hole mass is as large as $\sim$1.3 $\times$ 10$^{10}$ $M_{\odot}$. Therefore, these uncertainties and/or their combinations could lead to the total jet power and the accretion power exceeding the Eddington luminosity.

Curvature in the $\gamma$-ray spectrum is a characteristic property of powerful FSRQs \citep[see e.g.,][]{2010ApJ...710.1271A,2014MNRAS.441.3591H}. Interestingly, this feature is seen to be more prominent during the $\gamma$-ray flaring activities \citep[see e.g.,][for 3C 454.3 and 3C 279 respectively]{2011ApJ...733L..26A,2015arXiv150107363P}. During the period of $\gamma$-ray outburst, a significant curvature is noticed in the $\gamma$-ray spectrum of S5 0836+71 (Table~\ref{tab:gamma_spec}) similar to that reported by \citet{2013A&A...556A..71A}. A possible explanation of such curvature could be due to the attenuation of $\gamma$-rays by photon-photon pair production on He~{\sc ii} Lyman recombination lines within the BLR \citep{2010ApJ...717L.118P}. Moreover, \citet{2013ApJ...771L...4C} have proposed an alternative model in which a log parabolic particle energy distribution  and Klein-Nishina (KN) effect on IC scattering of BLR photons reproduce the observed curvature. Here, the spectral curvature is explained on the basis of KN mechanism and broken power law electron distribution. Therefore, though the high redshift and steep $\gamma$-ray spectrum suggests that S5 0836+71 would be weak in the $\gamma$-ray window of the SED, the giant GeV outburst of this source reveals many peculiar characteristics which are generally seen only in low redshift blazars.

\begin{figure*}
\hspace{-0.8cm}
\hbox{
      \includegraphics[width=9cm]{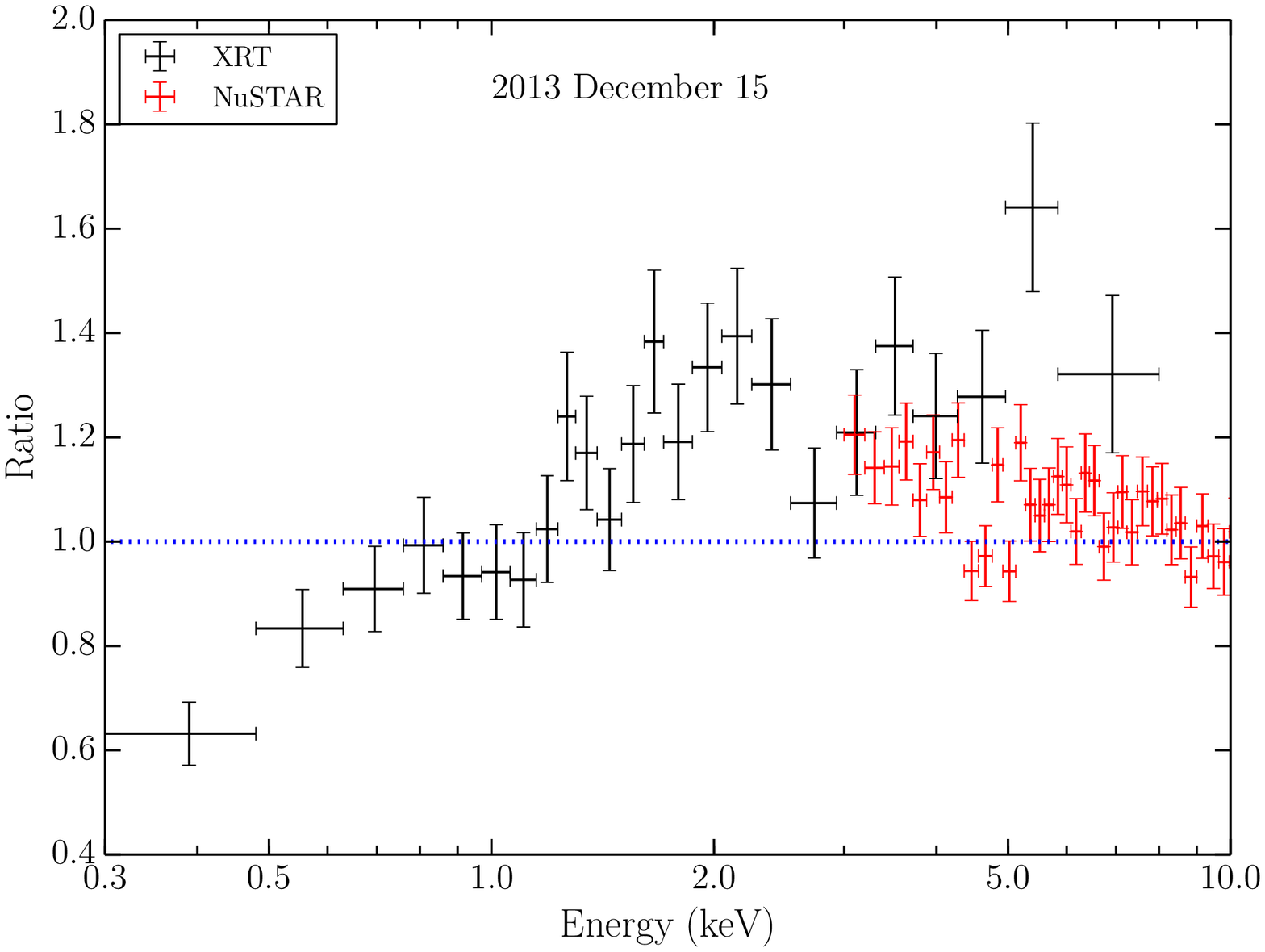}
      \includegraphics[width=9cm]{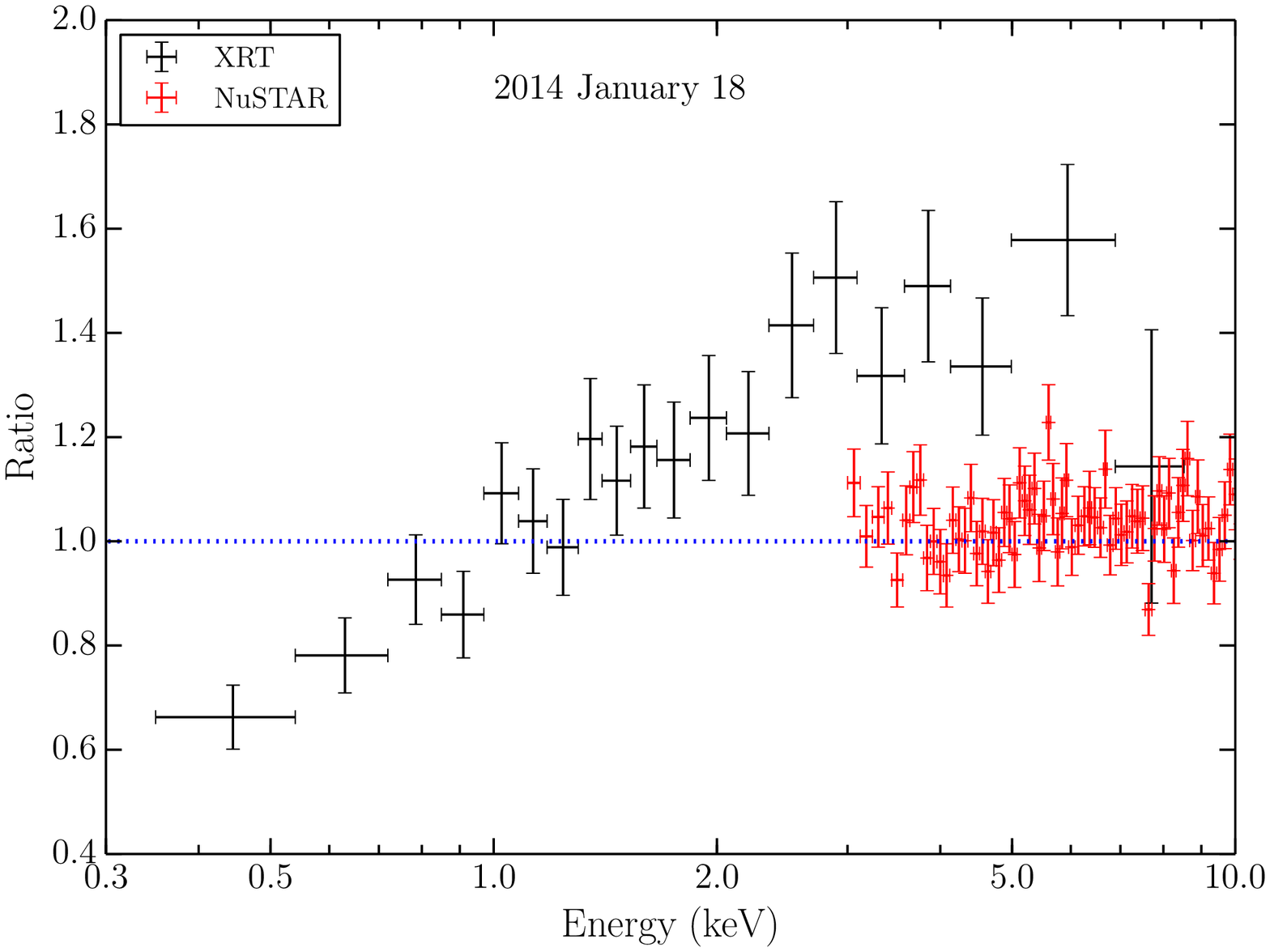}
     }
\caption{Data to model ratio of the fit of joint XRT and \nustar~spectra with a power law and fixed galactic absorption. The deficit of soft photons below 1 keV clearly indicates a greater degree of absorption or an intrinsic hardening of the spectrum.}\label{fig:ratio_abspl}

\end{figure*}

The data to model ratio of {\it Swift} observations, shown in Figure~\ref{fig:ratio_abspl} along with the \nustar~data, show a clear hardening of the X-ray spectrum below 1 keV, equally well fitted by a power law model with absorption in excess of the Galactic value or by a broken power law continuum with Galactic absorption. In fact, X-ray observations of high redshift quasars indicate for the presence of significant excess column densities in radio-loud sources \citep[e.g.,][and references therein]{2005MNRAS.364..195P,2007ApJ...665..980T}. Moreover, in several high redshift blazars the hardening of the soft X-ray spectrum is explained in terms of absorption by warm plasma ($N_{\rm H} \sim$10$^{22}$ cm$^{-2}$) surrounding the quasar, and probably in the form of outflow/wind \citep[e.g.,][]{1999MNRAS.308L..39F}. Since blazars typically do not show any significant intrinsic absorption, as the jet is expected to remove the gas from its vicinity, the presence of large gas column along the line of sight is highly improbable. One of the alternative explanation could be the underestimation of Galactic hydrogen column density by a factor of $\sim$3 obtained in this work, similar to that noted by \citet{2013A&A...556A..71A}. In the source frame, the Fe-K emission line is located at around 6.4 keV, and for S5 0836+71 it is expected to be around 2 keV in the observer's frame. From Figure~\ref{fig:ratio_abspl}, it can be seen that there is slight excess emission at $\sim$2-3 keV, however, due to short exposure of the XRT and hence low counting statistics, a strong claim cannot be made.

The homogeneous one zone synchrotron plus IC model with the inclusion of thermal emission from the accretion disk, successfully reproduces the observed SEDs. However, it does not explain the radio data, suggesting that much of the radio emission does not come from the compact emission region, primarily due to synchrotron self absorption process. The observed radio emission can be due to a superposition of many jet components \citep{1981ApJ...243..700K}, which is also supported by the lack of correlation seen in radio and $\gamma$-ray fluxes \citep{2013A&A...556A..71A}. The optical-UV part of the SEDs in all the activity states is dominated by the radiation from the accretion disk and this explains the lack of significant variability seen in the optical-UV band. Optical polarization observations during the peak of the $\gamma$-ray flare reveals the detection of very small polarization varying between 1\% to 6\% \citep{2013EPJWC..6104003J}, strengthening the idea of the dominance of optical-UV emission from the accretion disk. Further, the X-ray to high energy $\gamma$-ray window of the SEDs is explained via IC scattering of the photons originating outside the jet, primarily from the accretion disk and the BLR. Though in many FSRQs the origin of X-rays is attributed to SSC process \citep[e.g.,][]{2013ApJ...771L...4C}, it is better explained by EC mechanism for the case of high redshift blazars \citep{2010MNRAS.405..387G}. In this work, the X-ray spectrum is mainly reproduced by EC-disk and $\gamma$-ray emission via EC-BLR mechanism, as can be seen in Figure~\ref{fig:sed_model}.

\begin{figure*}
\hbox{
      \includegraphics[width=\columnwidth]{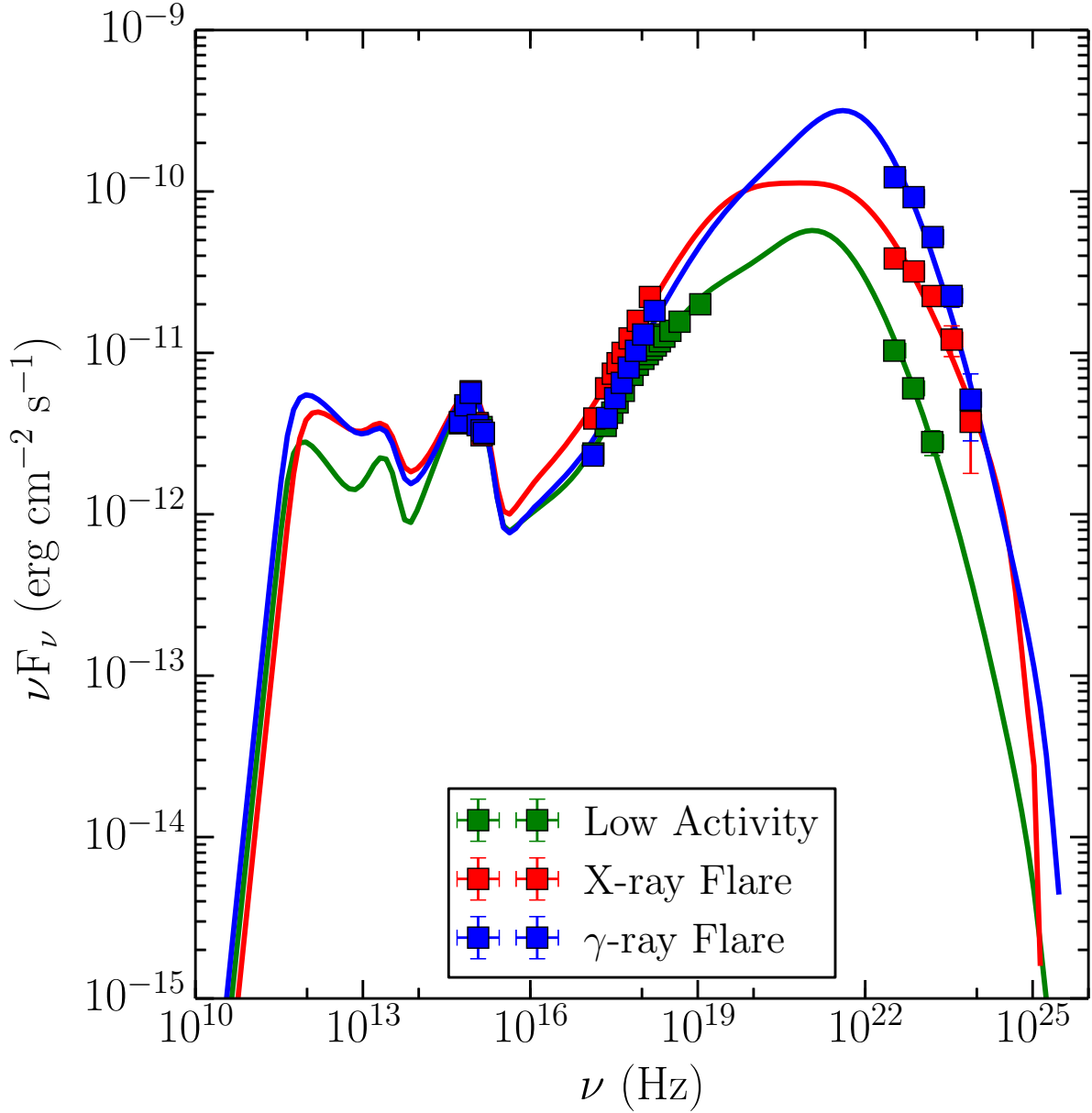}
     }
\caption{The SEDs of S5 0836+71 in different activity states, same as in Figure~\ref{fig:sed_model}, but plotted together for comparison.}\label{fig:sed_all}
\end{figure*}

To compare the physical properties of S5 0836+71 in different activity states, all the three SEDs are plotted together in Figure~\ref{fig:sed_all}. Looking at the obtained parameters in Table~\ref{tab:sed_par}, the emission region is found to be inside the BLR in all the activity states, similar to that found in earlier studies \citep{2007ApJ...669..884S,2007ApJ...665..980T,2008MNRAS.385..283C,2010MNRAS.405..387G}. Comparing the parameters obtained from the modeling of the low activity state SED with that obtained during the $\gamma$-ray flare suggests the enhancement of bulk Lorentz factor as a major cause of the flare. However, it should be noted that these results can also be reproduced by keeping the bulk Lorentz factor fixed and varying the total jet power and the equipartition condition. In such a situation, the required jet power would be less for the low activity state. Moreover, there are few other changes, such as slight hardening of the spectral shape of the particle energy distribution, location of the emission region (and hence its size), and decrease in the magnetic field. The Compton dominance (which is the ratio of IC to synchrotron peak luminosities) also increased by about an order of magnitude. Though the synchrotron flux also enhanced during the $\gamma$-ray flare, its contribution to the observed optical-UV spectrum is negligible and thus no major change in the optical band is noted. 

While the three SED models are close to equipartition between electrons and magnetic fields ($P_{\rm e}$ $\sim$ $P_{\rm B}$), the magnetic power is tiny compared to the proton power, or total jet power. This suggests a very low magnetization of the emitting region. Moreover, the radiative efficiency of the jet (expressed as $P_{\rm r}/P_{\rm jet}$), for F$_{\rm G}$, Q, and F$_{\rm X}$ periods, are obtained as 0.23, 0.03, and 0.20 respectively. This suggests that during the flares, a good fraction of the total jet power gets converted to the radiative power. However, majority of the jet power remains in the form of kinetic power and used to produce large scale jets. Looking at the various jet powers in Table~\ref{tab:sed_par}, it is clear that both $P_{\rm e}$ and $P_{\rm B}$ are smaller than $P_{\rm r}$. The fact that $P_{\rm e}$ $<$ $P_{\rm r}$ appears to be a contradiction at first sight since the radiation produced by electrons, in principle, cannot have more power than the emitting electrons. However, this inference may not be correct because the cooling timescale of $\gamma$-ray producing electrons are much shorter than $R_{\rm blob}/c$. Therefore, in order to maintain the emission over the entire flare period, the number of high energy electrons should be replenished till the flare duration. In other words, the electron population continuously gets energized at the expense of bulk energy and transfer the power to radiation. Hence $P_{\rm r}$ should be treated as the fraction of bulk luminosity transferred to radiation through electrons. On the other hand, since the mean energy of the electrons will be closer to $\gamma_{\rm min}$, $P_{\rm e}$ is mainly decided by the total number of electrons at lower energies which do not emit at $\gamma$-rays.

An interesting phenomenon is the detection of large X-ray flare with relatively moderate $\gamma$-ray variability amplitude (see period F$_{\rm X}$ in Figure~\ref{fig:mw_lc}). Since in this work, the X-ray emission is mainly explained by EC-disk mechanism, this suggest that a possible cause of the X-ray flare could be enhancement in the EC-disk flux. One of the possibility, perhaps, is the emission region to be located close to the central black hole. This is because the energy density of the direct disk radiation, as seen in the comoving frame, increases with decrease in the dissipation distance \citep[e.g.,][]{2009ApJ...692...32D,2009MNRAS.397..985G}. The variation of comoving frame energy densities as a function of the distance of the emission region from the central black hole for different activity periods, are shown in the bottom panels of Figure~\ref{fig:sed_model}. It can be seen in this plot that during the X-ray flaring period, the emission region is located in a jet environment where the accretion disk energy density dominates over other energy densities. Due to this effect, the enhancement of EC-disk radiation will be higher as compared to EC-BLR component and accordingly the rise in the X-ray flux will be more. 

\section{Summary}\label{sec:summary}
The detailed multi-frequency study of the high redshift blazar S5 0836+71 is presented in this work. The main findings are as follows
\begin{enumerate}
\item The long term multi-wavelength light curves of S5 0836+71 show more than one episodes of flaring activities in X-ray and $\gamma$-ray bands, while optical-UV fluxes show little or no variations.
\item The highest three hours binned $\gamma$-ray flux and associated photon index are found to be (5.22 $\pm$ 1.10) $\times$ 10$^{-6}$ \phflux~and 2.62 $\pm$ 0.27 respectively, measured on MJD 55866. This corresponds to an isotropic $\gamma$-ray luminosity of (1.62 $\pm$ 0.44) $\times$ 10$^{50}$ \lum, thus making it one of the most luminous $\gamma$-ray flare ever observed from the blazar class of AGN.
\item The observed $\gamma$-ray flux variability of about three hours is the shortest flux rising/decaying time ever measured from this source, and probably from any high redshift blazar beyond redshift 2.
\item A search for possible correlation between $\gamma$-ray flux and photon index, using a Monte Carlo simulation approach, does not result in any significant correlation between them.
\item Reproduction of the observed SEDs with a simple one zone leptonic emission model suggests the emission region to be located inside the BLR, and the optical-UV spectrum is dominated by the accretion disk radiation.
\item Enhancement in the bulk Lorentz factor is probably a primary factor of the observed giant $\gamma$-ray outburst. The high activity seen in the X-ray band with less variable $\gamma$-ray counterpart can be explained by locating the emission region closer to the central black hole where the comoving frame energy density from the accretion disk dominates over other energy densities, resulting in enhanced EC-disk flux peaking in X-rays. 
\end{enumerate}
\acknowledgments
The author thank the referee for constructive comments and C. S. Stalin and Sunder Sahayanathan for useful discussions. A substantial help in \nustar~data analysis received from Michael Parker is also acknowledged. This research has made use of data, software and/or web tools obtained from NASA’s High Energy Astrophysics Science Archive Research Center (HEASARC), a service of Goddard Space Flight Center and the Smithsonian Astrophysical Observatory. Part of this work is based on archival data, software, or online services provided by the ASI Science Data Center (ASDC). This research has made use of the XRT Data Analysis Software (XRTDAS) developed under the responsibility of the ASDC, Italy. This research has also made use of the NuSTAR Data Analysis Software (NuSTARDAS) jointly developed by the ASI Science Data Center (ASDC, Italy) and the California Institute of Technology (Caltech, USA). Use of {\it Hydra} cluster at the Indian Institute of Astrophysics is acknowledged.

\bibliographystyle{apj}
\bibliography{Master}

\begin{table}
\center
\caption{Fractional rms variability amplitude parameter ($F_{\rm var}$), calculated for the light curves shown in Figure~\ref{fig:mw_lc} and \ref{fig:nustar_lc}.}\label{tab:f_var}
 \begin{tabular}{lc}
\hline \hline
Energy band & $F_{\rm var}$\\
\hline
$\gamma$-ray (0.1$-$300 GeV) & 0.79 $\pm$ 0.02\\
X-ray (3$-$79 keV) & 0.21 $\pm$ 0.01 \\
X-ray (0.3$-$10 keV)  & 0.33 $\pm$ 0.01\\
UVW2 & 0.07 $\pm$ 0.02\\
UVM2 & 0.04 $\pm$ 0.05\\
UVW1 & 0.09 $\pm$ 0.02\\
U & 0.05 $\pm$ 0.02\\
B & 0.04 $\pm$ 0.03\\
V & 0.04 $\pm$ 0.08\\
\hline
\end{tabular}
\end{table}

\begin{table}
\center
\caption{Parameters obtained from the time profile fitting of two flares. Errors are estimated at 1$\sigma$ level.}\label{tab:flare_fit}
 \begin{tabular}{ccccccc}
\hline \hline
Name & $F_{\rm c}$ & $F_{\rm p}$ & $t_{\rm p}$ & $T_{\rm r}$ & $T_{\rm d}$ & $\chi^2_{\rm r}$\\
\hline
F1 & 1.11 $\pm$ 0.27 & 8.41 $\pm$ 2.01 & 55866.31 & 0.13 $\pm$ 0.05 & 0.10 $\pm$ 0.04 & 0.77\\
F2 & 1.31 $\pm$ 0.26 & 3.54 $\pm$ 1.29 & 55872.75 & 0.08 $\pm$ 0.06 & 0.27 $\pm$ 0.13 & 0.21\\
\hline
\end{tabular}
\tablecomments{Fluxes $F_{\rm c}$ and $F_{\rm p}$ are in 10$^{-6}$ \phflux, $T_{\rm r}$ and $T_{\rm d}$ are in days, and $t_{\rm p}$, which is kept fixed, is in MJD.}
\end{table}

\begin{table}\small
\caption{Results of the Model Fitting to the $\gamma$-ray Spectra of S5 0836+71, obtained for different time periods. Col.[1]: period of observation (MJD); Col.[2]: activity state; Col.[3]: model used (PL:
power law, LP: logParabola); Col.[4]: integrated $\gamma$-ray flux (0.1$-$300 GeV), in units of 10$^{-7}$
\phflux; Col.[5] and [6]: spectral parameters (see definitions in the text); Col.[7]: test statistic; Col.[8]: test statistic of the curvature.}\label{tab:gamma_spec}
\begin{center}
\begin{tabular}{cccccccc}
\hline\hline
Period & Activity & Model & $F_{0.1-300~{\rm GeV}}$ & $\Gamma_{0.1-300~{\rm GeV}}/\alpha$ & $\beta$ & TS & $TS_{\rm curve}$\\
~[1] & [2] & [3] & [4] & [5] & [6] & [7] & [8] \\
\hline
55,860$-$55,930 & $\gamma$-ray flare (F$_{\rm G}$) & PL & 6.25 $\pm$ 0.21 & 2.67 $\pm$ 0.04 &                & 2621.34& \\
                &                    & LP & 5.85 $\pm$ 0.22 & 2.50 $\pm$ 0.05 & 0.29 $\pm$ 0.05& 2638.81& 51.0\\
56,000$-$56,657 & Low activity (Q)   & PL & 0.51 $\pm$ 0.04 & 3.09 $\pm$ 0.10 &                & 242.03 & \\
                &                    & LP & 0.49 $\pm$ 0.04 & 3.04 $\pm$ 0.11 & 0.13 $\pm$ 0.12& 241.75 & 1.26\\
56,790$-$56,865 & X-ray flare (F$_{\rm X}$)& PL & 2.22 $\pm$ 0.15 & 2.59 $\pm$ 0.06 &                 & 582.70& \\
                &                    & LP & 2.05 $\pm$ 0.16 & 2.40 $\pm$ 0.10 & 0.18 $\pm$ 0.07& 586.40 & 8.96\\
\hline
\end{tabular}
\end{center}
\end{table}

\begin{table*}\scriptsize
\begin{center}
{
\caption{Summary of the Joint XRT+\nustar~Spectral Fitting. Col.[1]: fitted model (PL: power law, BPL: broken power law, GA: galactic absorption, AA: additional galactic absorption, zA: absorption in the intrinsic source frame); Col.[2]: additional neutral Hydrogen column density (10$^{20}$ cm$^{-2}$); Col.[3]: intercalibration constant ; Col.[4]: photon index of PL model or photon index before break energy in BPL model; Col.[5]: photon index after break energy in BPL model; Col.[6]: break energy (keV); Col.[7]: normalization at 1 keV (10$^{-3}$ ph cm$^{-2}$ s$^{-1}$ keV$^{-1}$); Col.[8]: statistical parameters.}\label{tab:xrt_nustar}
\begin{tabular}{lccccccc}
\hline\hline
 Model & $N^{\rm addi.}_{\rm H}$ & CONST & $\Gamma_{\rm X}$/$\Gamma_{\rm 1}$&  $\Gamma_{\rm 2}$& $E_{\rm break}$ & Norm. & $\chi^{2}$/dof.\\
 (1) & (2) & (3) & (4) & (5) & (6) & (7) & (8)\\
\tableline
 & & & & 2013 December 15 & & &\\
 PL+GA  & ...  & 0.81 $\pm$ 0.03 & 1.58 $\pm$ 0.01 & ... & ... & 3.02 $\pm$ 0.09 & 692.30/550\\
PL+GA+zA&87.25 $\pm$ 10.94&0.85 $\pm$ 0.03&1.64 $\pm$ 0.02&...&...&3.48 $\pm$ 0.12&573.44/549\\
PL+GA+AA &8.31 $\pm$ 0.93&0.88 $\pm$ 0.03&1.66 $\pm$ 0.02&...&...&3.57 $\pm$ 0.12&559.43/549\\
BPL+GA & ... &0.93 $\pm$ 0.06&1.13 $\pm$ 0.10&1.67 $\pm$ 0.03&2.07 $\pm$ 0.37&2.46 $\pm$ 0.18&553.30/548\\
\tableline
 & & & & 2014 January 18 & & &\\
 PL+GA  & ...  & 0.72 $\pm$ 0.02 & 1.66 $\pm$ 0.01 & ... & ... & 5.54 $\pm$ 0.13 & 880.74/762\\
PL+GA+zA&106.51 $\pm$ 13.84&0.83 $\pm$ 0.03&1.69 $\pm$ 0.01&...&...&5.90 $\pm$ 0.14&779.32/761\\
PL+GA+AA&9.58 $\pm$ 1.11&0.87 $\pm$ 0.03&1.69 $\pm$ 0.01&...&...&5.97 $\pm$ 0.14&766.55/761\\
BPL+GA& ... &0.98 $\pm$ 0.08&1.23 $\pm$ 0.10&1.70 $\pm$ 0.02&2.83 $\pm$ 0.62&3.71 $\pm$ 0.33&755.79/760\\
 \tableline
\end{tabular}
}
\end{center}
\end{table*}

\normalsize

\begin{table*}
\begin{center}
{
\small
\caption{Summary of the SED analysis. \fermi-LAT analysis results are given in Table~\ref{tab:gamma_spec}.}\label{tab:sed_flux}
\begin{tabular}{ccccccc}
\tableline\tableline
 & & & \nustar & & &\\
 Activity state & Exp.\tablenotemark{a} & $\Gamma_{3-79~{\rm keV}}$\tablenotemark{b} & $F_{3-79~{\rm keV}}$\tablenotemark{c} & Norm.\tablenotemark{d} & Stat.\tablenotemark{e} & \\
 \tableline
Q           & 29.7   & 1.64 $\pm$ 0.04 & 4.99$^{+0.19}_{-0.21}$ & 3.38 $\pm$ 0.28 & 425.79/436 & \\
\tableline
 & & & {\it Swift}-XRT  & & &\\
 Activity state & Exp.\tablenotemark{a} & $\Gamma_{0.3-10~{\rm keV}}$\tablenotemark{f} & $F_{0.3-10~{\rm keV}}$\tablenotemark{g} & Norm.\tablenotemark{d} & Stat.\tablenotemark{e} & \\
 \tableline
F$_{\rm G}$ & 20.65  & 1.21 $\pm$ 0.02 & 3.02$^{+0.07}_{-0.08}$ & 2.58 $\pm$ 0.06 & 265.68/313 & \\
Q           & 8.06   & 1.34 $\pm$ 0.04 & 2.36$^{+0.09}_{-0.10}$ & 2.35 $\pm$ 0.09 & 143.15/126 & \\
F$_{\rm X}$ & 2.32   & 1.26 $\pm$ 0.06 & 4.37$^{+0.28}_{-0.28}$ & 3.98 $\pm$ 0.24 & 80.88/56   & \\
\tableline
 & & & {\it Swift}-UVOT &  & & \\
 Activity state & V\tablenotemark{h} & B\tablenotemark{h} & U\tablenotemark{h} & UVW1\tablenotemark{h} & UVM2\tablenotemark{h} & UVW2\tablenotemark{h} \\
 \tableline
F$_{\rm G}$&3.73 $\pm$ 0.16&4.74 $\pm$ 0.14&5.66 $\pm$ 0.12&3.57 $\pm$ 0.05&3.31 $\pm$ 0.05&3.20 $\pm$ 0.03 \\
Q          &3.66 $\pm$ 0.20&4.72 $\pm$ 0.18&5.62 $\pm$ 0.18&3.60 $\pm$ 0.11&3.47 $\pm$ 0.06&3.16 $\pm$ 0.05 \\
F$_{\rm X}$&3.73 $\pm$ 0.33&4.48 $\pm$ 0.25&5.78 $\pm$ 0.18&3.68 $\pm$ 0.13&3.10 $\pm$ 0.10&3.16 $\pm$ 0.06 \\
\tableline
\end{tabular}
\tablenotetext{1}{Net exposure in kiloseconds.}
\tablenotetext{2}{Photon index of the power law model in 3$-$79 keV energy range.}
\tablenotetext{3}{Power law flux in 3$-$79 keV energy range, in units of 10$^{-11}$ \ergflux.}
\tablenotetext{4}{Normalization at 1 keV in 10$^{-3}$ \phflux~keV$^{-1}$.}
\tablenotetext{5}{Statistical parameters: $\chi^2$/dof.}
\tablenotetext{6}{Photon index of the absorbed power law model in 0.3$-$10 keV energy band.}
\tablenotetext{7}{Unabsorbed flux in units of 10$^{-11}$ \ergflux, in 0.3$-$10 keV energy range.}
\tablenotetext{8}{Average flux in {\it Swift} V, B, U, W1, M2, and W2 bands, in units of 10$^{-12}$ \ergflux.}
}
\end{center}
\end{table*}

\begin{table*}
{\small
\begin{center}
\caption{Summary of the parameters used/derived from the modeling of the SEDs in Figure~\ref{fig:sed_model}. Viewing angle is taken as 3$^{\circ}$ and the characteristic temperature of the torus as 900 K. For a disk luminosity of 2.25 $\times$ 10$^{47}$ erg s$^{-1}$ and black hole mass of 3 $\times$ 10$^9$ \Msun, the size of the BLR is $\sim$0.5 parsec (1700 $R_{\rm Sch}$).}\label{tab:sed_par}
\begin{tabular}{lccc}
\tableline
\tableline
                                                                &          & Activity state &         \\
Parameter                                                       & F$_{\rm G}$& Q            & F$_{\rm X}$ \\
\tableline
Slope of particle spectral index before break energy ($p$)      & 1.6        & 1.7          & 1.6         \\
Slope of particle spectral index after break energy ($q$)       & 4.0        & 4.2          & 3.8         \\
Magnetic field in Gauss ($B$)                                   & 1.7        & 1.8          & 2.5         \\
Equipartition factor$^*$ ($\eta_{\rm equi.}$)                   & 0.89      & 2.74        & 1.43        \\ 
Bulk Lorentz factor ($\Gamma$)                                  & 19         & 10           & 15          \\
Break Lorentz factor ($\gamma'_{\rm b}$)                        & 84         & 69           & 73          \\
Maximum Lorentz factor ($\gamma'_{\rm max}$)                    & 5e4        & 5e4          & 2e4         \\
Dissipation distance in parsec ($R_{\rm Sch}$) & 0.14 (500) & 0.11 (400)   & 0.07 (250) \\
\hline
Jet power in electrons in log scale ($P_{\rm e}$)               & 45.83      & 45.62        & 45.57       \\
Jet power in magnetic field in log scale ($P_{\rm B}$)          & 45.88      & 45.18        & 45.41       \\
Radiative jet power in log scale ($P_{\rm r}$)                  & 47.84      & 46.58        & 47.23       \\
Jet power in protons in log scale ($P_{\rm p}$)                 & 48.17      & 48.06        & 47.93       \\
\tableline
\end{tabular}
\tablecomments{$^*$Equipartition factor is the ratio of the particle to magnetic energy density ($\eta_{\rm equi.} = U'_{\rm e}/U'_{\rm B}$).}

\end{center}
}
\end{table*}

\newpage

\newpage

\newpage

\newpage

\newpage

\newpage

\newpage

\newpage

\newpage

\end{document}